# Quadratic magneto-optical Kerr effect spectroscopy: Polarization variation method for investigation of magnetic and magneto-optical anisotropies

V. Wohlrath,[1] Z. Sadeghi,[1] J. Kimák,[1] K. Hovořáková,[1] P. Kubaščík,[1] E. Schmoranzerová[1], L. Nádvorník,[1] F. Trojánek,[1] P. Němec,[1,*] and T. Ostatnický[1]

[1]*Faculty of Mathematics and Physics, Charles University, Ke Karlovu 3, 121 16 Prague 2, Czech Republic*

[*]Author to whom any correspondence should be addressed.

E-mail: petr.nemec@matfyz.cuni.cz.

**Abstract**

We present a method for a precise determination of magnetic anisotropy and anisotropy of quadratic magneto-optical response of thin films of ferromagnetic and ferrimagnetic materials. The method is based on measurements of a magneto-optical response for light close to the normal incidence on the sample with a fixed position. The measurement is performed for a set of orientations of an external magnetic field and a series of incident light linear polarizations beyond the standard *s* and *p* orientations. Based on the symmetry of the signal, we are able to separate the part of magneto-optical response that is even with respect to magnetization and, in turn, to exclude all non-magnetic contributions which come from imperfections of the experimental setup or from the sample itself. It is, therefore, possible to study the sample placed inside a cryostat: the polarization changes due to cryostat windows and possible strain-induced optical anisotropy of the sample are removed by the applied data processing. Thanks to this, we can perform measurements on low or elevated temperatures (from 15 to 800 K in our case), making it possible to study the behavior of magnetic materials in different magnetic phases and/or close to phase transitions. The applicability of this experimental technique was tested by measuring the low-temperature response of two samples of ferromagnetic semiconductor (Ga,Mn)As with a different Mn content at several wavelengths, which enabled us to deduce the magnetic and quadratic magneto-optical anisotropies in this material. In particular, we observed that the anisotropy of quadratic magneto-optical coefficients in (Ga,Mn)As is much weaker than that reported previously for other magnetic material systems.

# 1. Introduction

Thin magnetic films are the basic building blocks of all spintronic devices which are usually considered as the most promising successor of the current information-technology devices [1,2,3]. The vast majority of these spintronic devices is based on a vertical transfer of charge and spin through nanometer-thin layers in "sandwich-like" layered magnetic structures, where information is stored in a relative orientation of the adjacent magnetic layers [4,5]. In particular, spin-transfer-torque magnetic random-access memory [6] can serve as an example of a device based on this working principle that is commercially available nowadays. Therefore, a very important property of these films is their magnetic anisotropy which characterizes what the stable positions of magnetization in the film are and how difficult it is to change the magnetization orientation. While magnetic anisotropy of bulk magnetic materials can be determined by many conventional techniques (e.g., torque magnetometry), this task is considerably more difficult in the case of ultrathin films. One technique, which can be applied even for (sub)nanometer-thick films [7,8,9], is measurement of magneto-optical (MO) response, which is a magnetization-dependent change of linear polarization of incident light. Depending on the experimental geometry, there exist several MO effects, among which the measurement in reflection geometry using the magneto-optical Kerr effect (MOKE) is the most frequently applied for the investigation of magnetic thin films on nontransparent substrates [10,11]. Phenomenologically, MOKE is usually classified according to the position of a magnetization vector $M$ with respect to the sample surface and the orientation of the wave vector of the light as polar (PMOKE), longitudinal (LMOKE), and transversal (TMOKE) [11]. These effects, which are all linear in magnetization amplitude (LinMOKE), are usually the strongest MO effects, and, therefore, the most frequently used. However, there exist also MO effects that are quadratic with respect to magnetization amplitude (QMOKE) [11]. The contribution of these effects to the overall MO response is the strongest when the normally incident light is reflected from a magnetic material with an in-plane orientation of magnetization, which is typically the case for thin magnetic films due to demagnetizing field [10,11]. Consequently, QMOKE has to be taken into consideration even when LinMOKE is used for magnetometry of such films.

There exist several MOKE-based experimental characterization techniques which were used for quite distinct purposes in literature. In the method with rotating field of a constant magnitude [12,13] (named ROTMOKE by the authors of [12]), the magnetic anisotropy of a studied sample was evaluated from LMOKE measurement. Here, the major challenge is a separation of MO signals due to LMOKE and QMOKE, which is essential for a precise evaluation of the sample magnetic anisotropy [12]. This was achieved by using different symmetries of LMOKE and QMOKE with respect to the orientation of the *magnetization vector*, which was controlled by rotating external magnetic field with a constant amplitude. For this approach to be applicable, it was necessary to perform a very careful orientation of the sample magnetic easy (or hard) axis with respect to the plane of incidence (PoI) of light by rotating the sample (with a precision of ±0.5°) while LMOKE hysteresis loops were measured [12]. The separation of LMOKE and QMOKE was further elaborated in the "rotating field" method [7,14], where their different symmetry with respect to the applied *magnetic field* orientation was used. By repeating the measurement for several rotations of the studied sample, this technique enabled not only to deduce the magnetic anisotropy of the studied film but also to show that LMOKE is isotropic in the crystal orientation while QMOKE has both isotropic and anisotropic contributions [7,14]. Similarly, the anisotropy of QMOKE in cubic crystals can be studied by the "eight-directional" method [8,15]. This technique is based on MOKE measurement under eight different $M$ orientations for different sample rotations with respect to the light PoI using, typically, angles of light incidence from 5° to 45° [8]. This technique, however, does not enable one to evaluate the sample magnetic anisotropy. In fact, a strict magnetic saturation, which cannot be reached for a large magnetic anisotropy of studied samples, is the major limitation here. Another technique is the "vectorial MOKE" method [16,17], where the polarization rotation due to LMOKE is measured simultaneously with a reflectivity change due to TMOKE [16,17]. The main objective here is a simultaneous measurement of two magnetization components (with respect to PoI) and, consequently, the measurements of corresponding critical fields in the hysteresis loops for different sample rotations [16,17]. Overall, all these MOKE-based techniques use LMOKE as the main workhorse MO effect and involve a sample rotation in a plane perpendicular to PoI (or in a plane perpendicular to the incident light direction for the normal incidence of light). Therefore, these techniques were predominantly limited to room temperature [7,8,12,13,14,15,16]. The reason is, that the sample rotation in a standard magneto-optical cryostat is rather a difficult task because a good thermal contact and perpendicularity of the sample surface plane with respect to the rotation axis and PoI has to be achieved simultaneously in the cryostat [17]. In this paper, we show that the same experimental capabilities can be achieved using our new method that utilizes QMOKE solely. As this technique does not require the sample rotation – instead, we are using a rotation of the polarization plane of incident linearly polarized light – and because it is working also for the normal incidence of light, it is very easy to use it with a cryostat. To demonstrate its capabilities, we studied magnetic and magneto-optic anisotropies in diluted magnetic semiconductor (Ga,Mn)As, which is magnetically ordered only at cryogenic temperatures.

At present, there is no broadly accepted terminology for MO effects that are quadratic in magnetization and that exist in the reflection geometry at normal incidence with in-plane position of $\boldsymbol{M}$ [18]. This effect is sometimes called "reflection analogy to the Voigt effect" [18,19], orientational magneto-optical effect [20,21], or Hubert-Schäfer effect [22]. However, to be compatible with previous MOKE-based experimental characterization techniques [7,8,12,14], we will use the term QMOKE in this paper. In a typically studied metallic ferromagnetic films, like Fe [8] and Ni [23], QMOKE is significantly smaller than LinMOKE and, therefore, it was usually considered only as a nuisance that distorts the hysteresis loops [19] or field angle dependencies [12] measured using LMOKE. However, in the last 20 years QMOKE termed as "huge" or "giant" was reported in diluted magnetic semiconductor (Ga,Mn)As [24], $Co_2FeSi$ Heusler compound [25], and in magnetic semiconductor (Eu,Gd)O [26], where it is reaching values as large as 1 deg at certain wavelengths [26]. Growing interest in QMOKE was also inspired by the proposed concept of antiferromagnetic spintronics, where antiferromagnets play an active role in devices [27,28]. However, as antiferromagnets are insensitive to moderate magnetic fields, the portfolio of magnetic characterization techniques available for their experimental study is rather limited, especially in the case of thin films [29]. Importantly, QMOKE is present even in compensated collinear antiferromagnets where LinMOKE vanishes [29], which enables to perform time-resolved studies in antiferromagnets [30,31,32]. Quadratic MO effects can be also used for imaging magnetic domains [10,11,33], as demonstrated in ferromagnetic (Ga,Mn)As [34], antiferromagnetic NiO [35,36,37] and CoO [38,39], and also for magnetometry, as shown in yttrium-iron garnet (YIG) [9,40]. In fact, (Ga,Mn)As is a remarkably interesting material for QMOKE-based experiments [18,24,41]. For example, it enabled a direct measurement of the three-dimensional magnetization vector trajectory after the impact of a laser pulse [42], which in turn led to the discovery of optical spin-orbit torque effect [43].

In this paper, we describe a new experimental technique for the evaluation of magnetic anisotropy and anisotropy of QMOKE in thin magnetic films, which is based on measuring a QMOKE-induced MO response as a function of the external magnetic field direction for different orientations of incident light polarization plane. As this technique does not require a sample rotation and because it is working also for the normal incidence of light on the sample, it is fully compatible with cryostats. Our comprehensive analysis of the measured MO data enables to separate an optical anisotropy of the sample from the true magnetic anisotropy of the magnetic film, which is a Mn-concentration-dependent mixture of cubic and uniaxial anisotropy in the studied films of (Ga,Mn)As with a thickness of 20 nm. We also illustrate that the correctness of the anisotropy separation procedure can be verified experimentally if the experiment is performed at several wavelengths. Moreover, we show that, unlike in previous studies performed in Fe [8,44], Ni [23], $Fe_3O_4$ [45], Heusler compounds [25,46], and YIG [40], the MO response quadratic in magnetization is nearly isotropic with respect to the (Ga,Mn)As crystal orientation. This in turn justifies the applicability of QMOKE microscopy [34] and visualization of the three-dimensional magnetization vector trajectory [41] for magnetic characterization of (Ga,Mn)As.

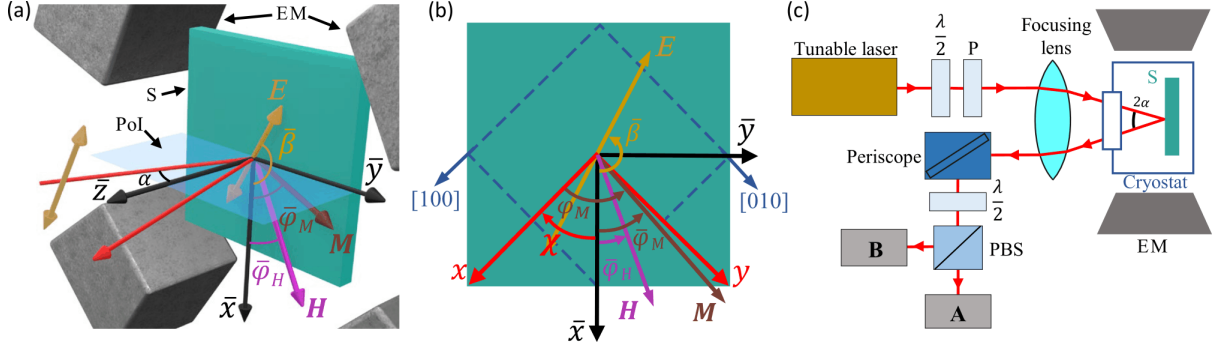

**Figure 1.** Schematic illustration of experimental geometry. (a) Light (red line) is incident at small angle ($\alpha$) on sample (*S*), which is placed in four-pole electromagnet (EM), defining the Plane of Incidence (PoI). Light polarization orientation (*E*), direction of magnetic field (*H*) and magnetization (**M**) are oriented at angles $\bar{\beta}$, $\bar{\varphi}_H$ and $\bar{\varphi}_M$, respectively, in the laboratory coordinate system ($\bar{x}, \bar{y}, \bar{z}$). (b) Depiction of the sample coordinate system ($x, y$), defined by crystallographic directions [100] and [010], which is rotated in the sample plane by angle $\chi$ with respect to the laboratory coordinate system. (c) Illustration of the experimental setup. Polarization orientation of incident light from tunable laser is set by half waveplate ($\lambda/2$) and polarizer (*P*). Light reflected from sample placed in cryostat is deflected by periscope, which is formed by two mirrors rotated with respect to each other that the overall optical anisotropy of the periscope is compensated. The polarization rotation is detected by polarization-bridge consisting of a balancing half waveplate ($\lambda/2$), polarization beam splitter (PBS) and pair of detectors (A and B). The angle of incidence $\alpha$ is strongly exaggerated for clarity; in reality, $\alpha \approx 1°$.

## 2. Theory

We aim at the determination of the magnetic and magneto-optical anisotropy of thin ferro- and ferrimagnetic films, using the measurement of QMOKE as the probe tool. Our goal is to be able to study samples at temperatures different from the ambient temperature, when the sample is placed in a cryostat at a fixed position. This light-polarization-based experiment is complicated by several factors. In addition to the exploited MO activity, the sample can also pose a magnetic field-independent optical anisotropy due to strain. Moreover, the cryostat windows typically show a nonzero MO response. As we demonstrate in this paper, the symmetry of the sample's QMOKE with respect to the incident light polarization and magnetization orientation allows one to separate QMOKE from other sources of the light polarization change. The only assumption in our analysis is that the quadratic MO response of all optical elements in the experimental setup is negligible compared to the QMOKE of the studied sample and that LinMOKE does not significantly exceed 10 mrad.

First, we define the coordinate systems in figures 1(a) and 1(b). There are two natural definitions, the first one being the *laboratory coordinate system* (overbar variables), the other one is the *sample coordinate system* which is related to the sample crystallographic axes (plain variables). These two systems are conjugate by a rotation by an in-plane angle $\chi$: $\bar{\varphi} = \varphi + \chi$. Although the laboratory coordinates appear naturally in the experiment, we will mainly use the sample coordinate system because the crystal lattice is the source of the anisotropies in which we are interested.

We start our analysis be revealing the QMOKE symmetry. At first, we consider a thin layer made from a fully isotropic medium except for its in-plane magnetization orientation in the direction defined by an in-plane angle $\bar{\varphi}_M$ which is measured from the axis $\bar{x}$ (here, we can use the laboratory coordinates due to the medium isotropy). From now on, we focus only on light at normal incidence in the reflection geometry. The optical response in reflection is proportional to the sample thickness if this is much less than the probe light wavelength. Under this assumption, it can be shown that the MO response reveals only the QMOKE contribution [see appendix A and equations (A20)-(A22) therein] which is expressed as:

$$\Delta\beta(\bar{\varphi}_M, \bar{\beta}) = P \sin 2(\bar{\varphi}_M - \bar{\beta}). \tag{1}$$

The symbol $\bar{\beta}$ denotes the orientation of the incoming light polarization, $\Delta\beta$ is the polarization rotation angle (that is typically less than few mrad) and $P$ is the QMOKE coefficient which depends on the permittivity tensor components. (Note that we are using such a sign convention that $P < 0$ corresponds to a *smaller* reflectivity for light polarized along the magnetization orientation than that for light polarized perpendicularly to it, see equation (2) in [41].) For the fixed $\bar{\varphi}_M$, the response can be rewritten as a linear combination of the goniometric functions $\sin 2\bar{\beta}$ and $\cos 2\bar{\beta}$:

$$\Delta\beta(\bar{\varphi}_M, \bar{\beta}) = P\sin 2\bar{\varphi}_M \cos 2\bar{\beta} - P\cos 2\bar{\varphi}_M \sin 2\bar{\beta}. \tag{2}$$

The proportionality of the result to the sum of goniometric functions of $2\bar{\beta}$ is a general result for any material under the assumption that the angle of incidence is zero. In real material, however, the multipliers of the goniometric functions do not obey the simple expression of equation (2) since the system symmetry is broken by the presence of the crystal lattice. We choose a (001)-oriented cubic crystal as an example; a similar calculation is possible, however, for any crystal symmetry [47]. We may derive (see appendix A) the following expression for a (Ga,Mn)As sample, which is in the sample coordinate system:

$$\Delta\beta(\varphi_M, \beta) = P_4 \sin 2\varphi_M \cos 2\beta - P_S \cos 2\varphi_M \sin 2\beta. \tag{3}$$

Here, two different QMOKE coefficients $P_4$ and $P_S$ arise as a consequence of the crystal-induced anisotropy and their ratio reflects the ratio of the magneto-optical $G$-tensor components [see equation (A21)]. We denote this effect as the *QMOKE anisotropy* throughout this paper. To understand this term from the experimental point of view, we first note that the magnetization makes the material birefringent[1] and thus equation (3) should follow the relation:

$$\Delta\beta(\varphi_M, \beta) = P(\varphi_M) \sin 2[\varphi_{\text{opt}}(\varphi_M) - \beta]. \tag{4}$$

The amplitude of the polarization rotation $P(\varphi_M)$ can be then denoted as the *(anisotropic) QMOKE amplitude* which depends on the magnetization orientation. The position of the birefringent axis is denoted as $\varphi_{\text{opt}}$ and it may be different from $\varphi_M$ as will be discussed in detail later. The above equation means that for a fixed orientation of the magnetization $\varphi_M$, polarization rotation is always a harmonic function of the incoming polarization angle $\beta$. However, contrary to the isotropic case described by equation (1), the polarization rotation angle depends on $\varphi_M$, *i.e.*, the QMOKE amplitude depends on $\varphi_M$. By expanding equation (4) and comparing it with equation (3), we find that $P(\varphi_M)$ and $\varphi_{\text{opt}}$ are unambiguously related to $P_4$, $P_S$ and $\varphi_M$. The square of the QMOKE amplitude can be expressed as:

$$P^2(\varphi_M) = (P_4 \sin 2\varphi_M)^2 + (P_S \cos 2\varphi_M)^2 = \frac{P_S^2 + P_4^2}{2} + \frac{P_S^2 - P_4^2}{2} \cos 4\varphi_M. \tag{5}$$

The QMOKE amplitude $P(\varphi_M)$ reveals the fourfold symmetry with extrema $P(0°) = P_S$ and $P(45°) = P_4$, respectively. To quantify the difference between $P_S$ and $P_4$, we define the relative QMOKE anisotropy as

$$\delta P = \frac{|P_S| - |P_4|}{|P_S| + |P_4|}. \tag{6}$$

The relative QMOKE anisotropy acquires the values from -1 to +1, thus reflecting which of the $P_S$ or $P_4$ terms are dominant, and $\delta P = 0$ if the QMOKE coefficient is independent of $\varphi_M$. The anisotropy of the QMOKE response has already been discussed in literature [7,8,14,15,23,44,45]. The anisotropy is widely defined in terms of the $G$-tensor components as $\Delta G = G_{1111} - G_{1122} - 2G_{2323} = G_S - 2G_4$ and this quantity is called MO anisotropy parameter [7,15,45] or anisotropic strength of $G$-tensor [8]. To avoid ambiguity, we will use the latter name here. The definition of the anisotropic strength of $G$ has a clear mathematical justification but it is not ideal from the experimental point of view because the value of $\Delta G$ alone does not tell if the QMOKE amplitude is isotropic or not. To justify this point, let us assume $P_S = P_4$. Using equation (A21) and the above definition of $\Delta G$, we get $\Delta G = 0$ and equation (5) gives the isotropic $P(\varphi_M)$. Assuming $P_S = -P_4$, on the other hand, we get large $\Delta G$ but, according to equation (5), the QMOKE amplitude is isotropic again. Therefore, our definition of the relative QMOKE anisotropy $\delta P$ by equation (6) better reflects the experimentally determined anisotropy of the magneto-optical activity due to QMOKE. (A more detailed discussion is provided in section 5.)

---

[1] The term *birefringent* means here that there are two optical eigenmodes coupled to the incoming field, whose wave vectors and/or attenuation coefficients differ. We use, therefore, the term *birefringent* as an abbreviation for the consequence of both the birefringence (i.e., different refractive indexes) and/or dichroism (i.e., different absorption coefficients).

In an experiment, the angle of the in-plane magnetization vector $\varphi_M$ in general differs from the orientation of the externally applied in-plane magnetic field, pointing at an angle $\varphi_H$. The reason is that the external field is not always strong enough to fully saturate the magnetization. In other words, magnetization is pushed by the local gradient of the magnetization free energy functional upon the change of the external field. The free energy functional for in-plane magnetization of a cubic, (001)-oriented sample with an additional uniaxial anisotropy reads:

$$\frac{F(\varphi_M)}{\mu_0 M_S} = -H\cos(\varphi_M - \varphi_H) + \frac{H_u}{2}\sin^2(\varphi_M - \varphi_u) + \frac{H_c}{2}\sin^2\varphi_M \cos^2\varphi_M \ . \tag{7}$$

Here, $H_c$ and $H_u$ are the cubic and uniaxial magnetic anisotropy constants, respectively, $H$ is the intensity of the external magnetic field and an angle $\varphi_u$ describes the orientation of the uniaxial anisotropy. For weak external fields, the local energy minimum does not coincide with the global energy minimum, *i.e.*, the magnetization orientation depends on its preceding position [see figure 8(a)]. For field intensity much higher than the coercive field, on the contrary, the magnetization always follows the global energy minimum, thus unambiguously defining the relation $\varphi_M(\varphi_H)$ [figure 8(b)]. Experimentally, the polarization rotation $\Delta\beta(\varphi_H, \beta)$ is always measured while the relation $\varphi_M(\varphi_H)$, denoted as the *magnetic anisotropy* in the following, is a result of the performed data analysis.

## 3. Experimental setup

The experimental setup is illustrated in figure 1(c). We used a supercontinuum laser (SuperK EXTREME, NKT Photonics) generating laser light in a very broad spectral range (400 – 2400 nm), from which the laser beam at desired wavelength can be selected by SuperK VARIA tunable filter (400 – 840 nm) or individual band pass filters, with a typical spectral width of 10 nm. The orientation of incident light linear polarization, i.e. a polarization angle $\bar{\beta}$, was set by a broadband half-wave plate (made by Newport: 10RP52-1B in spectral range 400-700 nm, 10RP52-2B in spectral range 700-1000 nm, and 10RP52-3B in spectral range 1000-1600 nm) and a polarizer (#47-316 Edmund Optics, for 400-700 nm and LPNI100, Thorlabs, for 650-2000 nm). Laser beam was focused by a lens on the surface of a sample, placed inside a cryostat (CS202PI-DMX-12, ARS; temperature range 15-800 K) surrounded by a custom-made computer-controlled four-pole vector magnet, which is similar to that described in [13] (see Supplementary material [48], sections S1 and S2 for more information). The sample was held at a temperature ≈ 15 K and its crystallographic axis [100] was rotated with respect to the laboratory axis $\bar{x}$ such that $\bar{\varphi}_H = \varphi_H + \chi$ with $\chi \approx -45°$. The angle of incidence $\alpha$ of the laser beam on the sample was set to ≈1° [see figure 1(a)], which is small enough to suppress LMOKE but significantly large to enable a spatial separation of the reflected beam from the incident one. Polarization of the reflected beam is in general elliptical with the orientation of the main polarization axis at the angle $\bar{\beta} + \Delta\beta$, where $\Delta\beta$ is the polarization rotation angle discussed in the previous section. We detected the polarization rotation using a Glan-Laser polarizer-based optical bridge (see appendix B of [49] for a detailed description) balanced by a $\lambda/2$ waveplate (always of the same type as that used at the laser output). The vector magnet was operated in two distinct modes. For the hysteresis loops measurements, magnetic field was applied along a selected in-plane crystallographic direction and its magnitude was changed. In the field rotation mode, field magnitude was set to a constant value (207 mT throughout this study) and its direction was rotated in the sample plane around the full 360° angle with a step 5°. After every change of the incident light polarization, the optical bridge has to be balanced [49], i.e., the half waveplate in front of PBS [see figure 1(c)] has to be rotated to a position where the difference between signals from detectors A and B is zero. The magnetic field may not be, however, strong enough to saturate the magnetization in an intended orientation and therefore we lack an absolute reference point. This is not a complication for our data evaluation method: each set of data for a given incident polarization and varying magnetic field is measured with the fixed balancing wave plate position in the optical bridge and the correct reference point for $\Delta\beta$ is fitted afterwards as described in the next section. To overcome the ambiguity of the unknown reference in the presentation of the experimental data in the field rotation mode, the curves of $\Delta\beta$ are centered around the value $\Delta\beta = 0$.

We used two 20-nm-thick $Ga_{1-x}Mn_xAs/GaAs$ samples with (001) sample plane orientation and with nominal Mn doping $x = 3\%$ and $x = 7\%$, which were prepared by molecular beam epitaxy using the optimized synthesis protocols [50]. The compressive in-plane stress together with demagnetizing field keeps the magnetization vector in the sample plane even if the sample plane is not perfectly aligned with the plane, where the orientation of the external magnetic field is changed. Further details about the studied samples are provided in [50]. The basic magnetic characterization of the studied samples can be performed by measuring QMOKE-based hysteresis loops, which are shown in figure 2 for the sample with Mn doping $x = 3\%$. The presence of M-shaped hysteresis loops [figures 2(a) and 2(d)] is a signature of the presence of four energetically equivalent magnetization easy axes in the sample plane [41,51,52]. In particular, the observed field-induced magnetization reversal is a combination of spin rotations and irreversible jumps between 4 in-plane magnetic domains, which exist in this material (see detailed discussion in [51]). As derived in [41], the amplitude *A* of the QMOKE-based hysteresis loops [see figure 2(a) for its definition] possess a characteristic harmonic dependence on the incident light polarization, which is indeed observed experimentally in our sample with *x* = 3% – see figures 2(b) and 2(e). Importantly, this amplitude is given not only by the QMOKE coefficient *P*, which is proportional to the square of the magnetization in-plane projection (see equations (5) – (8) in Supplementary information of [42]) but it depends also on the angle between two adjacent easy axes (see equation (11) in [41] and the adjacent discussion). If hysteresis loops are measured as a function of applied in-plane field direction, the angular dependence of the coercive fields $H_{c1}$ and $H_{c2}$, which are defined in figure 2(a), can be obtained – see figures 2(c) and 2(f). For completeness, we note that the correspondence of QMOKE-based hysteresis loops with those measured by magnetometry and by LMOKE can be found in [52] and [51], respectively. Moreover, the utilization of spatially-resolved images of QMOKE-based hysteresis loops for mapping of magnetization switching fields in GaMnAs-based microbars is demonstrated in [53]. The presence of M-shaped QMOKE-based hysteresis loops is limited not only to GaMnAs, but it was reported in many different magnetic materials [25,26,54,55].

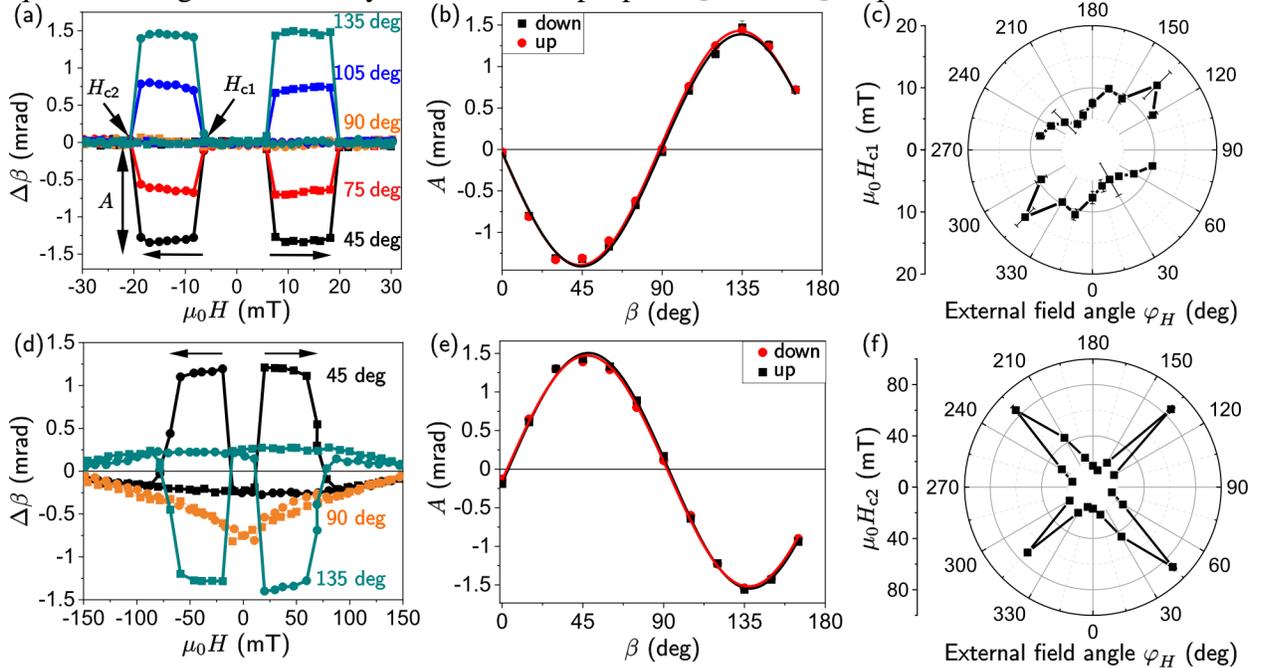

**Figure 2.** Characterization of sample with Mn doping $x = 3\%$ by QMOKE-based hysteresis loops measurements at 785 nm. Loops were measured for field directions $\varphi_H = 0$ deg (a) and 45 deg (d) using depicted orientations of incident light polarizations $\beta$. (b) and (e) Hysteresis loop amplitude $A$, defined in (a), as a function of $\beta$ deduced from data shown in (a) and (d), respectively (points). Lines are fits by equation (11) in [41] with $\xi = 135°$ (cf. figure 4 in the reference). (c) and (f) Dependence of coercive fields $H_{c1}$ and $H_{c2}$, defined in (a), as a function of magnetic field direction $\varphi_H$.

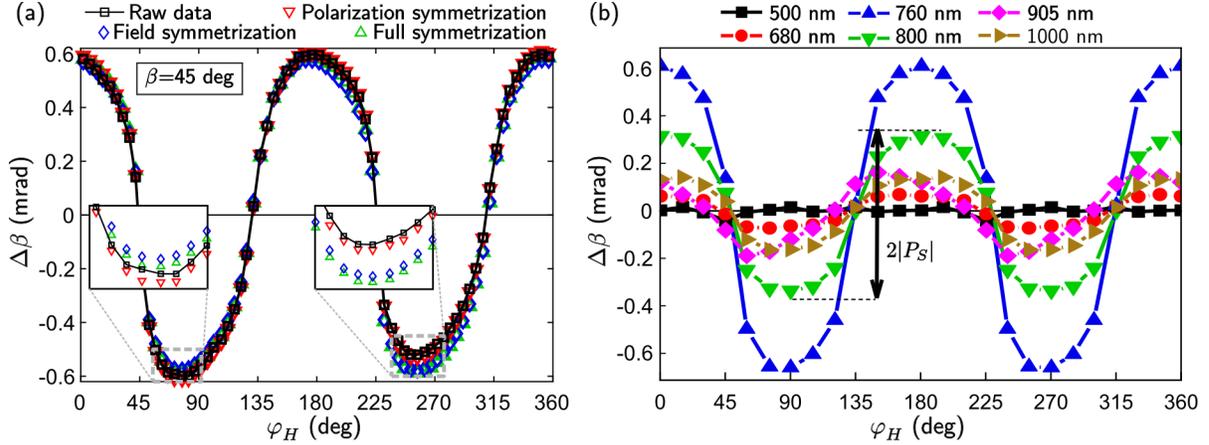

**Figure 3.** (a) Comparison of raw experimental data $\Delta\beta^{(\text{raw})}$ (black) measured at 760 nm for the sample $x = 3\%$ with results of symmetrizations. (b) Symmetrized curves $\Delta\beta^{(\text{sym})}(\varphi_H)$ for $\beta = 45°$ measured for a series of wavelengths.

## 4. Data processing and obtained results

The assumptions of the theory in the section 2 are not fully met in the experiment. In particular, we assumed zero angle of incidence, ideal optical components and their zero MO response. The experimental reality is, however, that the nonzero angle of incidence allows an admixture of the sample's LinMOKE to the studied QMOKE in the detected signal. Moreover, other optical components (cryostat windows, in particular) can contribute to the measured light polarization changes due to their LinMOKE. The used polarization optics can also exhibit some imperfections whose effect may be comparable to the measured signal, which is of the order of few hundreds µrad. It is, nevertheless, possible to significantly reduce the impact of the mentioned deviations from the ideal setup by symmetrization and fitting procedures. In this section, we first describe in part 4.1 how the raw experimental data are symmetrized in order to remove all LinMOKE contributions. In part 4.2, we describe a straightforward method for the measurement and evaluation of spectra of QMOKE coefficients. Nevertheless, as this method suffers from a relatively large experimental error, it is not suitable for a determination of the sample magnetic anisotropy. The procedure of the data acquisition and their processing in order to obtain the sample magnetic anisotropy and QMOKE coefficients with a much higher accuracy is described in part 4.3. Finally, the results of an evaluation of the QMOKE anisotropy are presented in Part 4.4.

### 4.1. Symmetrization

Our data analysis relies on the measured QMOKE signal, unlike previously reported methods based on LinMOKE [7,9,13,14,16,17,44,46]. The first step in the data analysis is therefore the removal of the LinMOKE contribution from the raw data $\Delta\beta^{(\text{raw})}(\varphi_H, \beta)$. This is done by a symmetrization of the data with respect to the orientation of the magnetic field $\varphi_H$: we can assume that the inversion of the external magnetic field direction (i.e., a shift by 180° in the measured data) causes inversion of the magnetization vector both in the sample and also in all other optical components (in particular cryostat windows). In such a case all LinMOKE contributions invert their signs while the QMOKE remains the same. Therefore, LinMOKE is removed and QMOKE conserved when averaging $\Delta\beta$ over $\varphi_H$ and $\varphi_H + 180°$. The second symmetrization is performed over the physically equivalent polarizations $\beta$ and $\beta + 180°$: even though it might seem useless, our experience indicates that it notably improves the measurement accuracy. The reason is, that the rotation of the incident polarization $\beta_1 \to \beta_2$, which is performed by rotating the $\lambda/2$ waveplate and polarizer [see figure 1(c)], must be accompanied by the rotation of the balancing $\lambda/2$ waveplate in the optical bridge by the angle $(\beta_2 - \beta_1)/2$. Polarization change by $180°$ therefore means rotation of the balancing waveplate by $90°$, which is not an equivalent position, and thus this averaging significantly reduces the influence of the balancing waveplate inaccuracy. In our data analysis, we apply both these symmetrizations on the raw experimental data $\Delta\beta^{(\text{raw})}$ in one step as follows:

$$\Delta\beta^{(\text{sym})}(\varphi_H, \beta) = \tfrac{1}{4}[\Delta\beta^{(\text{raw})}(\varphi_H, \beta) + \Delta\beta^{(\text{raw})}(\varphi_H + 180°, \beta)$$
$$+ \Delta\beta^{(\text{raw})}(\varphi_H, \beta + 180°) + \Delta\beta^{(\text{raw})}(\varphi_H + 180°, \beta + 180°)], \qquad (8)$$

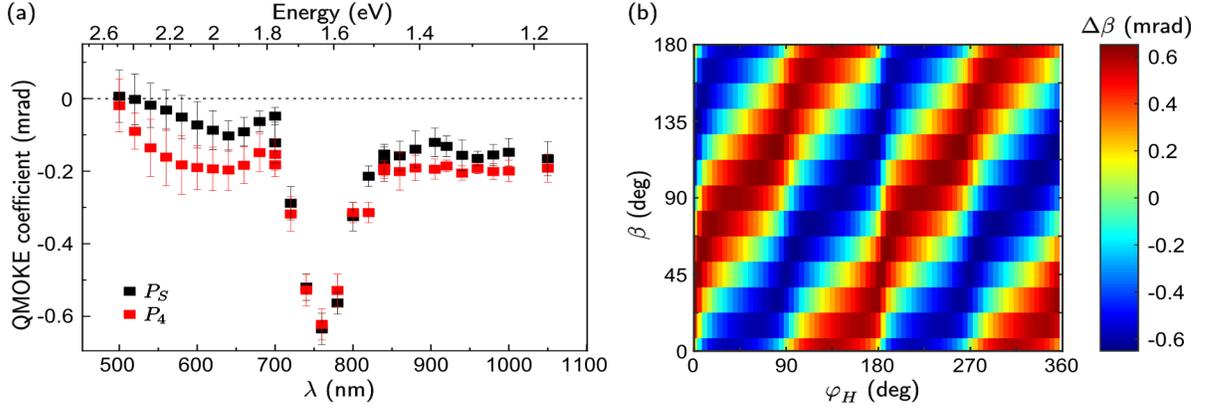

**Figure 4.** (a) Spectral dependency of QMOKE coefficients $P_4$ (red) and $P_S$ (black), obtained by the procedure described in section 4.2 for sample $x = 3\%$. (b) Colormap of the full set of symmetrized data $\Delta\beta^{(\text{sym})}(\varphi_H, \beta)$ measured at $\lambda=760$ nm.

As an example, we show in figure 3(a) the results of the two individual symmetrizations and the full symmetrization by equation (8) for the sample $x = 3\%$, probe wavelength 760 nm and incoming polarization angle $\beta = 45°$. In reality, however, even the symmetrized data are not an ideal QMOKE-based signal, described by equation (3). They contain all the terms which are even in magnetization amplitude, *i.e.,* in particular magnetization-independent contributions, QMOKE and their admixture. Magnetization-independent response appears naturally due to various symmetry breakings, mainly the non-zero angle of incidence, sample optical anisotropy or inaccurately balanced optical bridge. In our analysis, the non-QMOKE components are supposed to be at most comparable to the QMOKE magnitude. In that case, it is possible to deduce the QMOKE amplitudes (in particular, their spectral dependence) with a relatively high accuracy as described in the next paragraph. To deduce magnetic anisotropy from these experiments, however, it is necessary to use a more advanced method which is described later on.

## 4.2. QMOKE spectra

The QMOKE response is fully described by two sample-related parameters $P_4$ and $P_S$ which can be easily measured if we properly set the measurement geometry, namely the mutual orientation of the polarization angle $\beta$ and the external field orientation angle $\varphi_H$. It is clear from equation (3) that for example the $P_S$ coefficient is $P_S = -\Delta\beta(\varphi_M = 0°, \beta = 45°)$. However, due to unknown magnetic anisotropy, the magnetization orientation cannot be set to the required value of $\varphi_M$ simply by selecting $\varphi_H = \varphi_M$. Instead, we set $\beta = 45°$ and we rotated the magnetization around the full 360° angle span by rotating the external field direction. In such a case, $\Delta\beta^{(\text{sym})}(\varphi_M, 45°) \propto -P_S \cos 2\varphi_M$ with extrema appearing at $\varphi_M$ being multiples of 90°. The MO signal change between the maximum and the minimum in the $\varphi_M$-dependence is then $2|P_S|$. Although the angle $\varphi_M$ is not known, the measured QMOKE signal is a periodic function of the external field orientation $\varphi_H$ [see figure 3(b)] with the same span $2|P_S|$ between the extrema. Similarly, the other QMOKE coefficient $|P_4|$ can be measured by setting $\beta = 0°$ and repeating the procedure. The proper sign of the coefficient $P_S$ can be set as an opposite sign with respect to the datapoint $\Delta\beta^{(\text{sym})}(\varphi_H = 0°, \beta = 45°)$. For the $P_4$ coefficient, we set the same sign as $\Delta\beta^{(\text{sym})}(\varphi_H = 45°, \beta = 0°)$. The major advantage of this kind of measurement is that it is relatively fast because one needs to probe only four polarizations (including the polarizations shifted by 180° necessary for the symmetrization) at each wavelength and the data processing is quite simple. The resulting spectral dependency of the QMOKE coefficients for the $x = 3\%$ sample are plotted in figure 4(a). If a higher precision is needed, measurements at $\beta = 0°$ and 45° can be supplemented by measurements at $\beta = 90°$ and 135° (plus, again, at the polarizations shifted by 180° for the symmetrization), where the same shape of $\Delta\beta^{(\text{sym})}$ but with an opposite sign [e.g., $\Delta\beta^{(\text{sym})}(\varphi_M, 0°) = -\Delta\beta^{(\text{sym})}(\varphi_M, 90°)$] is expected [see equation (3)]. The examples of such data are discussed later in section 4.3 and shown in figure 6.

## 4.3. Analysis of magnetic anisotropy

Besides information about the QMOKE coefficients, the method of rotating field allows one to also deduce information about the position of the magnetization $\varphi_M(\varphi_H)$, depending on the orientation of the external field. We can easily identify the situations when the angle between magnetization and light polarization direction is an integer multiple of 45°, i.e., when the curves $\Delta\beta^{(\text{sym})}(\varphi_H, \beta)$ with $\beta = 0°$ and 45° discussed above reveal an extremum. If the external field is stronger than the sample coercive field [see figure 2(f)], the magnetization deflection angle $\varphi_M - \varphi_H$ is only several degrees. So, the closest multiple of 45° to $\varphi_H$, where the extreme appears, is the position of the magnetization. In principle, we can identify in this way several points of the mapping $\varphi_H \leftrightarrow \varphi_M$ and the rest is done by assuming that the mapped dependencies $\Delta\beta^{(\text{sym})}(\varphi_M, \beta)$ are, according to equation (3), a cosine or a sine of $2\varphi_M$. The obtained result is sensitive, however, to every misalignment, in particular to a setting of the polarization angle of incoming beam with respect to the crystallographic axes of the sample. To overcome this problem, we measured the polarization rotation $\Delta\beta^{(\text{raw})}(\varphi_H, \beta)$ for a set of polarization orientations $\beta$ around the whole 360° circle with the step of 15°. For each of these polarizations, we performed the measurement in the rotating magnetic field. Then we performed the symmetrization to get $\Delta\beta^{(\text{sym})}$, according to equation (8), with the example of the results depicted in figure 4(b). The last step is an extraction of the dependencies of the terms proportional to $\cos 2\beta$ and $\sin 2\beta$ on $\varphi_H$, as described in detail in appendix B. In short, we have to get rid of non-QMOKE contributions as discussed in the section 4.1. We performed a global fit of all the experimental data $\Delta\beta^{(\text{sym})}$ measured at various $\beta$ [see figure 4(b)] by equations (B2) and (B5) what allows us to rewrite equation (3) using the fitted dependencies:

$$\Delta\beta(\varphi_H, \beta) = \Delta\beta_4^{(\text{fit})}(\varphi_H) \cos 2\beta - \Delta\beta_S^{(\text{fit})}(\varphi_H) \sin 2\beta \ . \tag{9}$$

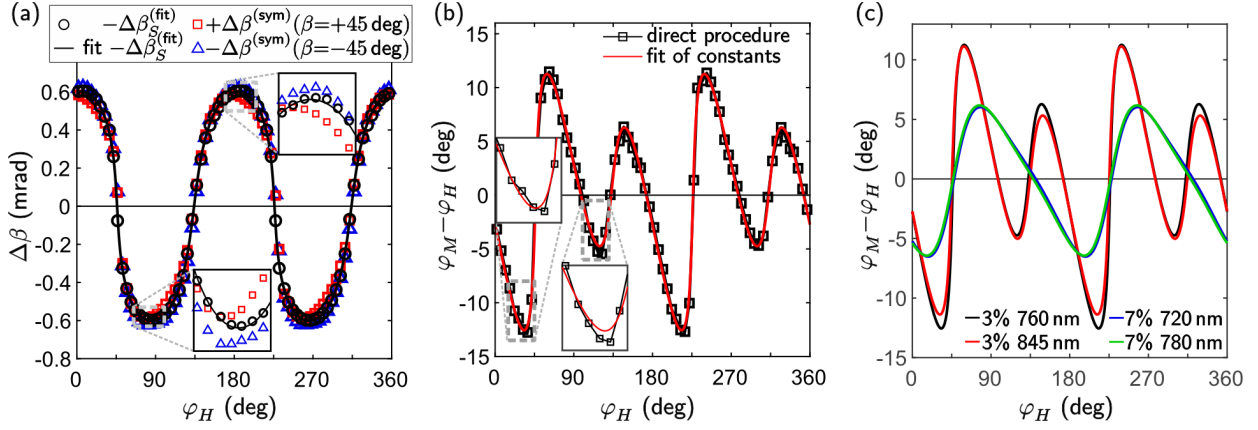

**Figure 5.** (a) Comparison of symmetrized data $\pm\Delta\beta^{(\text{sym})}(\varphi_H,\beta)$ for angles $\beta = \pm 45°$ (closed symbols) and fitted dependency $\Delta\beta_S^{(\text{fit})}(\varphi_H)$ (open circles) for $x = 3\%$ sample at 760 nm. The solid line shows theoretical curve $\Delta\beta(\varphi_H, 45°)$ with $\varphi_H \leftrightarrow \varphi_M$ mapping calculated using fitted magnetic anisotropic constants, acquired by the *fit of constants* procedure. (b) Deflection of magnetization with respect to external field $\varphi_M - \varphi_H$ analyzed by *direct procedure* (black) and by *fit of constants* procedure (red). (c) Magnetization deflection angle for both samples and two wavelengths for each sample as a result of the *fit of constants* procedure.

The fitting functions $\Delta\beta_4^{(\text{fit})}(\varphi_H)$ and $\Delta\beta_S^{(\text{fit})}(\varphi_H)$ here are in fact the traces $\pm\Delta\beta^{(\text{sym})}(\varphi_H, \beta)$ for $\beta = 0° (+), 45°(-)$ discussed in the previous paragraph from which the non-QMOKE artifacts were removed. This is shown in figure 5(a) where, as an example, we plot the obtained dependency $\Delta\beta_S^{(\text{fit})}(\varphi_H)$ (open circles) together with the symmetrized data for two incident polarizations $\pm\Delta\beta^{(\text{sym})}(\varphi_H, \beta = \pm 45°)$ (closed symbols). The difference between the data obtained by fitting and symmetrization is the accuracy of the results: the fit averages over more sets of data, it does not require precise adjustment of the polarization angle, and it removes experimental artifacts. To illustrate this, we plot data series for both samples and different wavelengths in figure 6. In both samples, fitted data sets agree very well with the symmetrized data when the measurements are performed at the QMOKE maximum for the particular sample, see figures 6(a) and 6(c). However, when MO data are measured outside this spectral range, where the MO response is weaker, the deviations between the individual curves start to be more pronounced, see figures 6(b) and 6(d). Consequently, it is essential to use $\Delta\beta_4^{(\text{fit})}(\varphi_H)$ [or at least the average of $\Delta\beta^{(\text{sym})}(\varphi_H, 0°)$ and $-\Delta\beta^{(\text{sym})}(\varphi_H, 90°)$] and $\Delta\beta_S^{(\text{fit})}(\varphi_H)$ [or the average of $\Delta\beta^{(\text{sym})}(\varphi_H, -45°)$ and $-\Delta\beta^{(\text{sym})}(\varphi_H, 45°)$] for any further analysis.

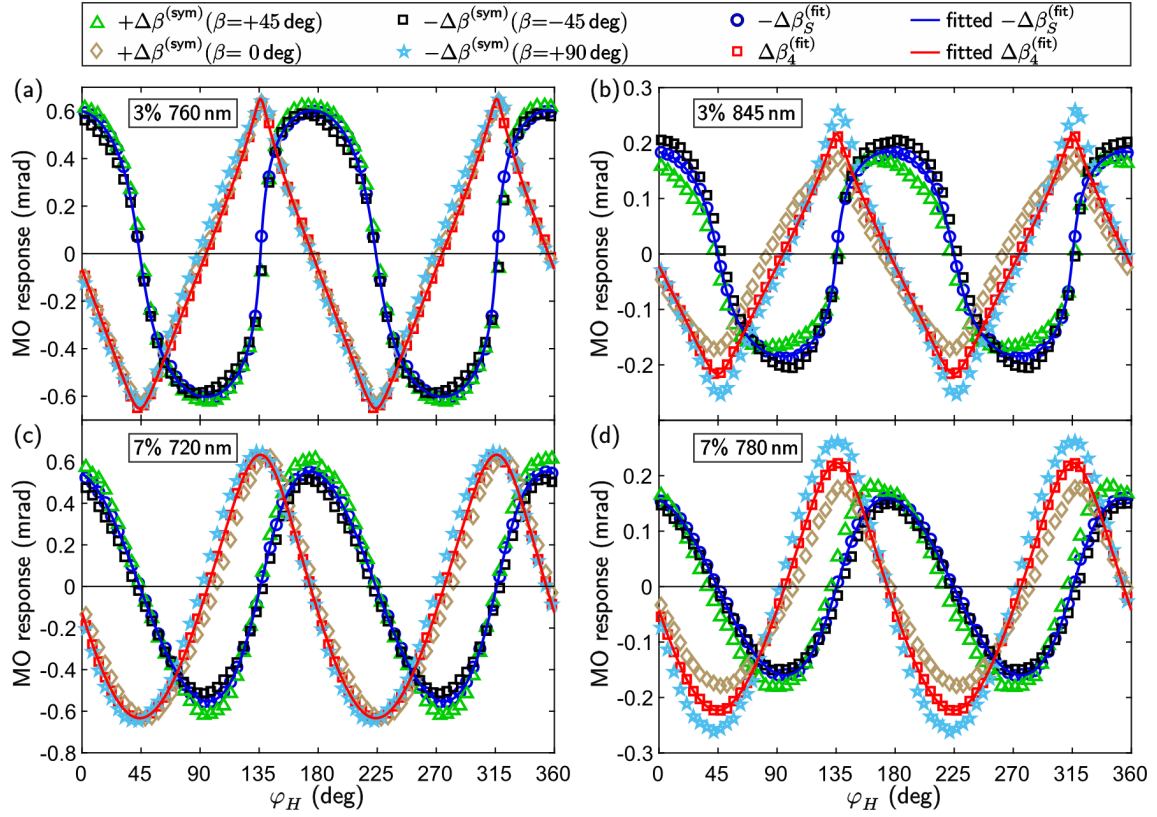

**Figure 6.** Comparison of symmetrized data $\Delta\beta^{(\text{sym})}(\varphi_H)$, fitted dependencies $\Delta\beta_4^{(\text{fit})}(\varphi_H), \Delta\beta_S^{(\text{fit})}(\varphi_H)$ and their numerical simulation, based on the values of table 1 in the main text. Results are plotted for the sample $x = 3\%$, 760 nm (a) and 845 nm (b), and sample $x = 7\%$, 720 nm (c) and 780 nm (d).

Finally, the two dependencies obtained by the fitting should depend on the magnetization orientation as $\Delta\beta_4^{(\text{fit})}(\varphi_H) = P_4 \sin 2\varphi_M$ and $\Delta\beta_S^{(\text{fit})}(\varphi_H) = P_S \cos 2\varphi_M$, respectively, which relate $\varphi_H \leftrightarrow \varphi_M$ and define the QMOKE coefficients $P_4$ and $P_S$. The actual evaluation of the magnetic anisotropy and QMOKE coefficients from these fitted dependencies can be found numerically using a rich variety of procedures, with two of them described in appendix B. In the first one (labelled as a *direct procedure* in the following), we adjust the QMOKE coefficients $P_4$ and $P_S$ fitting equations (B2) and (B5) and then we find a proper $\varphi_M$ from the ratio:

$$\frac{\Delta\beta_4^{(\text{fit})}(\varphi_H)}{\Delta\beta_S^{(\text{fit})}(\varphi_H)} = \left(\frac{P_4}{P_S}\right) \tan 2\varphi_M \qquad (10)$$

for each $\varphi_H$. The second and, in our opinion, more appropriate procedure (labelled as a *fit of constants*) is based on an assumption of particular magnetic anisotropies deduced from the known sample symmetry. Here, we fit the corresponding anisotropy constants to best reproduce equations (B3) and (B4) — in our case, we consider equation (7) for the magnetization free energy. Results of both procedures are shown in figure 5(b) where we display the resulting magnetization deflection angle $\varphi_M - \varphi_H$ derived for the $x = 3\%$ sample. Experimentally, the credibility of the obtained $\varphi_M - \varphi_H$ dependencies (i.e., their uncertainties) can be deduced if this procedure is performed at more than one wavelength of probing light because the magnetic anisotropy, which is the sample's property, has to be independent of the wavelength used. As we depict in figure 5(c) and table 1, in both studied samples the deduced magnetic anisotropies do not depend significantly on the measurement wavelengths, which is a confirmation of the reliability of the whole data measurement and analysis.

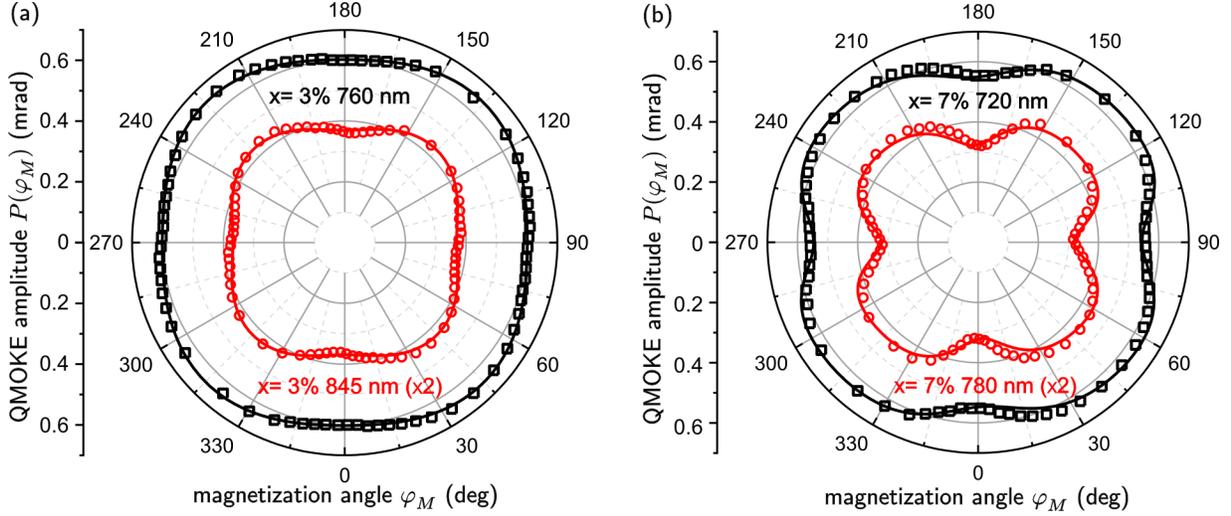

**Figure 7.** Extracted datapoints of the anisotropic QMOKE amplitude $P(\varphi_M)$ (symbols) and its fit (lines) using $P_4$, $P_S$ coefficients listed in table 1, for both samples $x = 3\%, 7\%$ and different wavelengths.

### 4.4. MOKE anisotropy

We may describe the QMOKE anisotropy in terms of coefficients $P_4$ and $P_S$ using equation (3) or, equivalently, as a general optical anisotropy (by using the anisotropic permittivity tensor) which depends on the orientation of the sample's magnetization. In our particular geometry, the optical anisotropy is described by two refractive indices of sample's linearly polarized (and mutually perpendicular) eigenmodes and an orientation $\varphi_{\text{opt}}$ of the eigenmode with the higher refractive index. The QMOKE response under these assumptions is expressed by equation (4) where a difference of the two refractive indices is combined into just one $\varphi_M$-dependent anisotropic QMOKE amplitude $P(\varphi_M)$. The dependence of this amplitude on $\varphi_M$ can be easily measured experimentally: for a fixed $\varphi_M$, we measured the polarization dependence of the QMOKE polarization rotation $\Delta\beta$ which should be a harmonic function of $2\beta$. The magnitude of $P(\varphi_M)$ is then [*cf.* equation (5)]:

$$|P(\varphi_M)| = \left[\left(\Delta\beta_4^{(fit)}(\varphi_M)\right)^2 + \left(\Delta\beta_S^{(fit)}(\varphi_M)\right)^2\right]^{\frac{1}{2}}. \quad (11)$$

The sign of $P(\varphi_M)$ depends on the choice of the angle $\varphi_{\text{opt}}$ which is not unique. Our sign convention is that $\varphi_{\text{opt}} = 0$ for $\varphi_M = 0$ and the sign of $P(\varphi_M)$ is set such that $P(\varphi_M = 0) = P_S$ [*cf.* equations (3) and (4)]. Obtained results for both studied samples and the used wavelengths are plotted in figure 7 as symbols. Theoretical fits, calculated using the parameters from table 1 and assuming the magnetization rotation in the external field with its free energy described by equation (7), are plotted as lines. Particularly good agreement between the measured and expected theoretical dependencies confirms not only the accuracy of our measurements but also the correctness of all the assumptions, in particular the one of the crystal symmetry in the studied samples.

| $x$ | $\lambda$ (nm) | $\mu_0 H_c$ (mT) | $\mu_0 H_u$ (mT) | $\varphi_u$ (deg) | $P_4$ (mrad) | $P_S$ (mrad) | $P_j$ error (mrad) | $P_4^*$ (mrad) | $P_S^*$ (mrad) | $P_j^*$ error (mrad) |
|---|---|---|---|---|---|---|---|---|---|---|
| 3% | 760 nm | 124 | 34 | 51 | −0.65 | −0.60 | 0.03 | -0.63 | -0.59 | 0.05 |
| 3% | 845 nm | 119 | 33 | 47 | −0.22 | −0.18 | 0.03 | -0.25 | -0.21 | 0.05 |
| 7% | 720 nm | 14 | 44 | 50 | −0.63 | −0.55 | 0.03 | -0.65 | -0.52 | 0.06 |
| 7% | 780 nm | 17 | 43 | 47 | −0.22 | −0.16 | 0.03 | -0.26 | -0.15 | 0.05 |

TABLE 1: Magnetic (columns 3-5) and QMOKE (columns 6-7) anisotropy parameters evaluated from the QMOKE data for both samples and two distinct measurement wavelengths. QMOKE coefficient error in column 8 is determined from the residuum of the fit. Columns 9-10, marked with an asterisk, are QMOKE coefficients evaluated directly from the symmetrized data, as described in section 4.2. The last column displays the estimate of the error of the QMOKE coefficients extracted from the symmetrized data.

## 5. Discussion

In this section, we discuss the obtained results, and we compare them to the previously published data. Although the experimental part of this study is focused on $Ga_{1-x}Mn_xAs$ crystals with the zinc-blende crystal structure, the theory and data processing method is applicable to any crystal with the same symmetry. For other symmetry groups, the permittivity tensor (A1) would have a different but still well defined form [47] and therefore a method of data processing similar to ours could be derived, following the steps presented in section 2 and appendix A. A more detailed discussion of the data fitting, according to appendix B, can be found in the Supplementary material [48], section S3.

Apart from being compatible with measurements in a cryostat, the advantage of the presented experimental technique is that it enables us to evaluate QMOKE coefficients in studied samples with a user-selectable precision and, therefore, time requirements. If we require only identification of a spectral range where the QMOKE response is relatively strong in a particular sample, it is sufficient to measure the data $\Delta\beta^{(\text{raw})}(\varphi_H, \beta)$ at experimentally available wavelengths for one input polarization $\beta$. Importantly, this measurement, which in unknown materials should precede the QMOKE-based magnetometry [56,57,58] or hysteresis loops measurement [24,26,41], is very fast. If the experiment is performed for two polarizations $\beta$ and $\beta + 45°$, also a rough estimate of the QMOKE coefficient anisotropy can be obtained, as illustrated in figure 4(a). Note that the precision of the achieved results is increased remarkably if symmetrized data $\Delta\beta^{(\text{sym})}(\varphi_H, \beta)$ are used instead of the as-measured data $\Delta\beta^{(\text{raw})}(\varphi_H, \beta)$. Moreover, further precision enhancement can be obtained if the average of "$\Delta\beta^{(\text{sym})}(\varphi_H, \beta)$" and "$-\Delta\beta^{(\text{sym})}(\varphi_H, \beta + 90)$" is always used for the selected $\beta$ – see figure 6. If detailed measurements for many polarizations [as in figure 4(b)] at several wavelengths are performed, the magneto-optical (figure 7) and magnetic [figure 5(c)] anisotropies can be studied in detail. Moreover, the spectral independence of the latter can serve as an extremely useful test of the results reliability.

To demonstrate the applicability of this technique, we studied samples of a ferromagnetic semiconductor $Ga_{1-x}Mn_xAs$ where the Curie temperature is well below the room temperature, i.e., the experiment has to be performed in a cryostat. Also, the magnetic [50] and magneto-optical [41] properties in this material system are known to be strongly dependent on the Mn content $x$. As a starting point, we discuss the deduced magnetic anisotropy in the studied samples. We obtained, using an average value from the measurements at different wavelengths (table 1), $\mu_0 H_c = (122 \pm 5)$ mT and $\mu_0 H_u = (34 \pm 5)$ mT for the $x = 3\%$ sample, and $\mu_0 H_c = (16 \pm 5)$ mT and $\mu_0 H_u = (44 \pm 5)$ mT for the 7% sample. Moreover, the orientation of the uniaxial anisotropy was identified to be the same in both samples, at $\varphi_u = (49 \pm 2)°$ (i.e., along the $[\bar{1}10]$ crystallographic direction when a suspected unintended samples misalignment on the cryostat cold finger of 4° is considered). The deduced values of anisotropy constants and the uniaxial anisotropy orientation are in a very good agreement with the results obtained in similar samples in [50] (see figure 4 in there), which were derived from a much more time demanding combination of SQUID and optical FMR experiments performed with magnetic fields applied at several sample orientations. (Note the difference between $\mu_0 H_j$ reported here and $K_j$ in refererence by a factor of two, which comes from a different definition of the anisotropy constants in our equation (7) and equation (S5) in the cited reference.) In particular, we observed a rather pronounced decrease of $H_c$ and a mild increase of $H_u$ with $x$, while $\varphi_u$ was independent on $x$.

We can estimate that the magnetization vector orientation is determined with an error of ~1°, taking into account the relative error of the polarization rotation measurement less than 5% [*cf.* closed symbols in figure 5(a)] and an error of the orientation of the incident light polarization below 0.5°. The error of the orientation of the external magnetic field is ~2° and its variation upon rotation around the full circle is ~5 mT, according to section S2 of the Supplementary material. We therefore suggest that in our setup, the magnetic field is the main source of the uncertainty and we put the values 5 mT and 2° as the errors of the anisotropy constants and $\varphi_u$, respectively.

Here, it is also appropriate to discuss the necessary strength of the rotating magnetic field for which our method is functional. For example, a considerable drawback of the eight-directional method [8,15] is a requirement to reach a full magnetization saturation by the applied field. In our method, on the other hand, we do not have such a requirement — we just need to be able to rotate the magnetization around the full circle when the driving field direction is changing and, at the same time, the magnetization must be homogeneous throughout the optically probed area of the sample. We therefore need to apply a field stronger than the maximum coercive field, which is $\approx 90$ mT in the studied samples of (Ga,Mn)As (see figure 2). In accord with this, we used a field magnitude of 207 mT in our experiments. In order to be able to determine the magnetic anisotropy coefficients, there is an additional limitation for the external field: it must not exceed certain value such that we are able to observe sufficient magnetization deflection angle $\varphi_M - \varphi_H$ that has to be larger than the measurement precission. Taking into account angular precission 2° and an estimate of the maximum deflection $\varphi_M - \varphi_H \approx H/H_u$ (or $H/H_c$ for the cubic anisotropy), we find the upper limit for the external field ~400 mT with the help of table 1.

We checked the sensitivity of the experimental results on the external field size by measuring the deflection angle in the $x = 3\%$ sample using $\mu_0 H$=50 mT and 207 mT and rotating the field in both directions, see figure 8. Apparently, when the field of 50 mT was used, the signal was strongly hysteretic. This effect was most pronounced around the hard axes positions, which are located at $\varphi_M \approx 45°$ plus multiples of $90°$, where the largest coercive field $H_{c2} \approx 90$ mT is observed, as shown in figure 2(f). In this case, the sample was probably in a multi-domain state (see figure 3 in [34]). In contrast, when a stronger field of 207 mT was used, the hysteretic behavior nearly disappeared, as expected for a single–domain state.

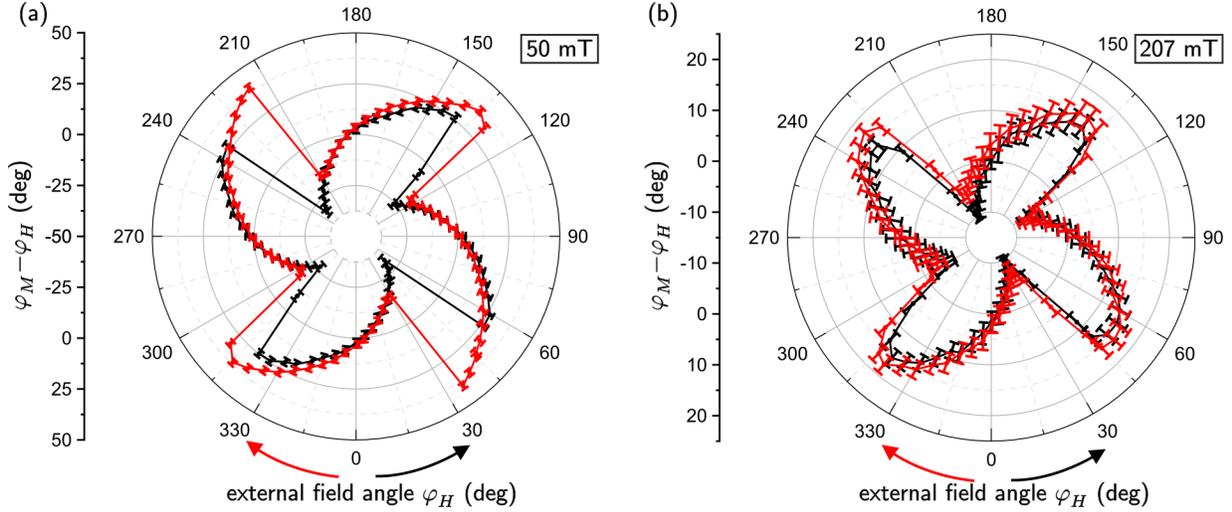

**Figure 8.** Magnetization deflection angle $\varphi_M - \varphi_H$ as a function of $\varphi_H$ for sample $x = 3\%$ measured using clockwise and countrer-clockwise rotation of field $\mu_0 H = 50$ mT (a) and 207 mT (b).

Figure 8 also illustrates why deducing the QMOKE coefficients from the measured MO data after symmetrization only is, in general, not very precise. To deduce the coefficient $P_4$ from the symmetrized data, the $\Delta\beta^{(sym)}(\varphi_H, \beta = 0°)$ curve has to be measured (see figure 6). If the applied field is not sufficiently strong, the experimentally measured extremal MO signals are *not observed* for the magnetization located at the hard axis $\varphi_M = 45°$, as could be expected from equation (3), but at a considerably different value of $\varphi_M$. For example, in the $x = 3\%$ sample and applied field of 50 mT the magnetization deflection angle can be as large as $35°$ [see figure 8(a)]. Consequently, the coefficient $P_4$ deduced from the symmetrized data would be $\cos 35° \approx 82\%$ of its correct value. In contrast, missing points around hard axes in dependencies $\Delta\beta_{4,S}^{(fit)}(\varphi_H)$ are not so critical if we apply the data fitting method. In particular, we checked that the results do not change significantly if we miss 30% of raw data in the vicinity of the hard axes.

Next, we discuss the obtained QMOKE magneto-optical coefficients. Up to know, spectral dependences of QMOKE coefficients in $\text{Ga}_{1-x}\text{Mn}_x\text{As}$ samples with a different $x$ were reported only in [18]. In that paper, the QMOKE amplitudes were deduced from a rather complex experiment where a custom-made rotating sample holder attached to the cryostat cold finger was used and the measurement procedure involved several steps where sample and magnetization rotations (to and from the easy axes) were combined. Moreover, as the plausible QMOKE anisotropy was not considered in the reference, we have to convert the MO coefficients reported therein to our notation before comparing the obtained results. For the sample $x = 3\%$, the cubic anisotropy dominates over the uniaxial one and, consequently, the easy axes are rather close to the main crystallographic directions [100] and [010]. Consequently, in [18] the measurement was performed with $\beta \approx 0°$ and, therefore, the reported QMOKE amplitude corresponds to the MO coefficient labelled here as $P_4$. The published value at 773 nm (i.e., close to the QMOKE maximum in this sample, see figure 3 in the reference) was $P_4 = -0.60 \pm 0.03$ mrad that is in accord with our value $P_4 = -0.65 \pm 0.03$ mrad at 760 nm (table 1). In the case of the sample $x = 7\%$, the easy axis is close to the [$\bar{1}10$] direction and, therefore, $\beta = 45°$ was used and the amplitude reported in [18] corresponds to $P_S$. The published value at 720 nm (at the QMOKE maximum) was $P_S = -0.44 \pm 0.02$ mrad while our value is $P_S = -0.55 \pm 0.03$ mrad. These values are rather similar but do not agree within the estimated errors. In a detailed discussion in section S3 of the Supplementary material [48], we have identified the possible source of this experimental error, which has been carefully removed in this study but that was not considered in [18] previously.

The observed anisotropy of QMOKE is rather weak in a maximum of the QMOKE spectral dependence and it increases for other wavelengths. Moreover, this anisotropy is increasing with the Mn-doping (*cf.* figure 7). Overall, the QMOKE anisotropy is considerably weaker in (Ga,Mn)As than in all other previously studied ferromagnetic materials, as illustrated in table 2 where we summarized previously reported results, which we converted to our notation for the sake of a quantification and a mutual comparison (see appendix A for details). We may speculate that this anisotropy suppression is a consequence of a relatively simple band structure of (Ga,Mn)As: The dominant contribution to the optical response comes from the (vertical) band-to-band transitions which are resonant with the optical field. This condition selects well defined states in the valence bands which contribute to the MO response and the QMOKE anisotropy then comes from the anisotropy of the model Hamiltonian with respect to the orientation of the magnetization. In (Ga,Mn)As the Hamiltonian (described by Eq (3) in [18]) contains terms that originate in *s-d* and *p-d* couplings, which are isotropic, while the anisotropy comes from the spin-orbit part. It can be shown that the kinetic exchange *p-d* coupling dominates over the spin-orbit interaction for the states of interest and, therefore, we expect mostly isotropic QMOKE in (Ga,Mn)As, which is indeed observed experimentally.

| Material | t (nm) | T (K) | λ (nm) | α (deg) | $|P_4|$ (mrad) | $|P_S|$ (mrad) | δP | source |
|---|---|---|---|---|---|---|---|---|
| $Ga_{0.97}Mn_{0.03}As$ | 20 | 15 K | 760 | 1 | 0.65 | 0.60 | -0.04 | this work |
| $Ga_{0.97}Mn_{0.03}As$ | 20 | 15 K | 845 | 1 | 0.22 | 0.18 | -0.10 | this work |
| $Ga_{0.93}Mn_{0.07}As$ | 20 | 15 K | 720 | 1 | 0.63 | 0.55 | -0.07 | this work |
| $Ga_{0.93}Mn_{0.07}As$ | 20 | 15 K | 780 | 1 | 0.22 | 0.16 | -0.16 | this work |
| $Co_2FeSi$ | 21 | room | 670 | 0.5 | 0.21 | 0.49 | +0.40 | figure 3 in [25] |
| $Co_2FeSi$ | 11 | room | 670 | 0.5 | 0.13 | 0.38 | +0.46 | figure 3 in [59] |
| $Co_2Mn_{0.77}Ge_{0.42}$ | 50 | room | 670 | 0 | 0.02 | 0.04 | +0.21 | figure 5 in [60] |
| $Co_2FeAl_{0.5}Si_{0.5}$ | 30 | room | 670 | 0 | 0.03 | 0.21 | +0.72 | figure 8 in [56] |
| $Co_2MnSi$ | 30 | room | 638 | 0 | 0.05 | 0.12 | +0.40 | figure 3 in [61] |
| $Co_2MnSi$ | 90 | room | 638 | 0 | 0.03 | 0.10 | +0.52 | figure 5(a) in [62] |
| $Co_2MnSi$ | 90 | room | 638 | 45 | 0.02* | 0.09* | +0.60* | figure 5(b) in [62] |
| $Co_2Mn_{1.30}Si_{0.84}$ | 30 | room | 632.8 | 45 | 0.04* | 0.13* | +0.53* | figure 5 in [46] |
| Fe | 50 | room | 670 | 3 | 0.07 | 0.02 | -0.47 | figure 2 in [15] |
| Fe | 12.5 | room | 670 | 45 | 0.06* | 0.34* | +0.69* | figure 6 in [8] |
| Ni | 20 | room | 635 | 45 | 0.01* | 0.02* | +0.23* | figure 1(c) in [23] |

TABLE 2: Comparison of magnitudes of magneto-optical coefficients $|P_4|$ and $|P_S|$, and the resulting QMOKE anisotropy δP for studied GaMnAs films with values reported for various ferromagnetic films (with thickness $t$, measured at temperature $T$, using light wavelength $\lambda$, and angle of incidence $\alpha$). The depicted values were deduced from reported graphs in listed references using the procedure described in appendix A. Results of the experiments performed at $\alpha$ significantly different from 0 deg, for which the theoretical formulas in appendix A were derived, are marked with "*" for clarity.

Another important aspect, which was not addressed in previous studies, is that the orientation of magnetization, $\varphi_M$, and that of the birefringent axis due to QMOKE, $\varphi_{\text{opt}}$, *do not* coincide in general. Let us first consider a fully isotropic QMOKE, *i.e.*, $P_4 = P_S$ (i.e., $G$-tensor components $2G_{2323} = G_{1111} - G_{1122}$). In this case, equation (3) simplifies to equation (1) and, by comparing it with equation (4), we find $P(\varphi_M) = const.$ and $\varphi_{\text{opt}} = \varphi_M$. When a large QMOKE anisotropy is introduced by setting $P_4 \neq 0$ and $P_S = 0$, equation (3) can be rewritten such that $\Delta\beta = P_4 \sin 2\varphi_M \sin 2(45° - \beta)$. Now, the $\varphi_M$-dependent anisotropic QMOKE amplitude is $P(\varphi_M) = P_4 \sin 2\varphi_M$ and the birefringent axis is located at a fixed orientation $\varphi_{\text{opt}} = 45°$. Further, for $P_4 = -P_S$, equation (3) gives $\Delta\beta = -P_4 \sin 2(-\varphi_M - \beta)$ and thus $P(\varphi_M) = -P_4$ and $\varphi_{\text{opt}}(\varphi_M) = -\varphi_M$. So, the birefringent axis can co-rotate with $\varphi_M$, counter-rotate or stay fixed, depending on the QMOKE anisotropy. More generally, if $P_4 \neq P_S$ (i.e., when QMOKE is anisotropic), $\varphi_{\text{opt}} \neq \varphi_M$ and, unlike the magnetization orientation, the dependency $\varphi_{\text{opt}}(\varphi_H)$ starts to be wavelength-dependent. As we show in figure 9, exactly this behavior is observed experimentally in the studied (Ga,Mn)As samples. In the sample $x = 3\%$ studied at 760 nm, when the QMOKE anisotropy is rather small (see table 2), positions of $\varphi_{\text{opt}}$ and $\varphi_M$ are very similar. When the detection wavelength is changed to 845 nm, which leads to a QMOKE anisotropy increase, the deviations between $\varphi_{\text{opt}}$ and $\varphi_M$ increase [see figure 9(a)]. Similar effect is apparent also in the sample $x = 7\%$ [figure 9(b)] but considerably more pronounced, due to its larger QMOKE anisotropy (*cf*. table 2).

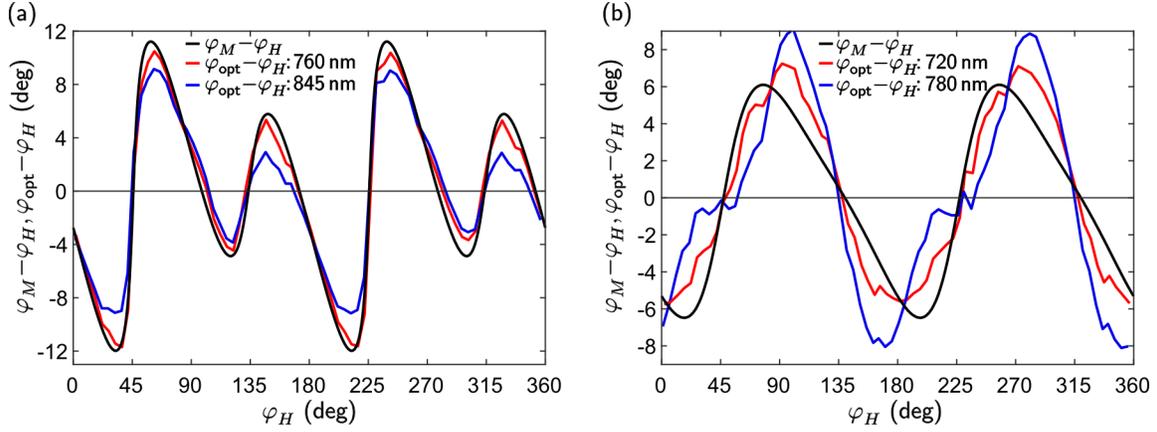

**Figure 9.** Comparison of deflection angles with respect to orientation of external field: wavelength-independent deflection of magnetization $\varphi_M - \varphi_H$ (black) and wavelength-dependent deflection of optical birefringent axis $\varphi_{opt} - \varphi_H$ (red, blue) measured in samples with $x = 3\%$ (a) and $7\%$ (b) with external field magnitude 207 mT.

Finally, we address the role of the symmetry of studied samples in this measurement technique. Up to know, we have analyzed and discussed the results measured in a material with a cubic crystal symmetry, thus considering a particular form of the $G$-tensor. If a sample symmetry were not known, it would be still possible to measure the QMOKE amplitude in a sense that $\Delta\beta^{(sym)}(\varphi_H, \beta)$ in some regular grid of parameters $\varphi_H$ and $\beta$ would be measured using the rotating field. As discussed in appendix A, polarization rotation $\Delta\beta^{(sym)}$ is a consequence of the sample birefringence and it is a time reversal-symmetric part of the overall MOKE due to the symmetrization. It is, therefore, a linear combination of $\cos 2\beta$ and $\sin 2\beta$ at each $\varphi_H$ for an arbitrary crystal symmetry. Consequently, the fit using equation (B2) is quite general and it removes unintentional offsets of the measured curves and also artifacts having the $4\beta$ arguments. By this way, it could be determined what is the QMOKE amplitude for a particular *orientation of the external field* $\varphi_H$ but this is not what equation (4) defines as the anisotropic QMOKE amplitude $P(\varphi_M)$, which is a function of the *orientation of the magnetization*. In particular, for other than cubic symmetry, the function $\Delta\beta_4^{(fit)}(\varphi_M)$ is not expected to be proportional to $\sin 2\varphi_M$ and, therefore, the fitted curves $\Delta\beta_{4,S}^{(fit)}$ would need to be processed in another way than we described here. Overall, there would be no way to obtain any relations between the permittivity tensor components if the crystal symmetry and orientation were not known. Similarly, as discussed in the previous paragraph, the orientation of the birefringent axis $\varphi_{opt}$ is directly measured in this experiment. The mutual orientation of $\varphi_{opt}$ and $\varphi_M$ depends strongly on the QMOKE anisotropy and consequently, as discussed above, information about the crystal symmetry is needed also for the determination of the magnetic anisotropy, *i.e.*, for the mapping $\varphi_{opt} \to \varphi_M$.

## 6. Conclusions

We present a contactless all-optical method for a determination of magnetic anisotropy and anisotropy of quadratic magneto-optical coupling constants in a ferro- or ferrimagnetic materials with an in-plane orientation of the magnetization and a cubic crystal symmetry. Unlike in previously reported studies where LinMOKE was used to determine the magnetic anisotropy constants [7,12,13,14,46], our method relies on the measurement of the QMOKE response and allows to simultaneously extract the magnetic and magneto-optical anisotropies. The advantage of the use of QMOKE as a probe is that normal incidence of light on the sample can be used. Moreover, if the signal symmetrization with respect to the magnetic field and incident polarization orientation is performed, only the sample contributes to the signal. Furthermore, there is no need to rotate the sample during the experiment because the incident light polarization plane is rotated instead. Finally, the full magnetization saturation by external magnetic field [8,15,23,25] in a preselected direction does not have to be achieved for this method to be applicable. Altogether, these features allow us to study samples placed in a cryostat, where we can measure at elevated or low temperatures, and thus extending the experimental possibilities towards observation of the material behavior at different magnetic phases or in the vicinity of phase transitions.

Based on symmetry considerations in the studied samples with a cubic symmetry, we derived the general formula [equation (3)] for the sample-induced polarization rotation dependency on the incoming light polarization $\beta$ and magnetization orientation $\varphi_M$, under the assumption of normal light incidence on a semi-infinite or an optically thin sample. It turns out that the response should follow the harmonic dependency with respect to $2\beta$ for a fixed $2\varphi_M$ and vice versa. We showed that from the experimental point of view, it is sufficient to measure the polarization rotation in a regular grid of external magnetic field orientations and incoming light polarizations. Such measurement can be automated and thus the whole measurement procedure does not exceed 1 hour at 1 wavelength. The data analysis involves symmetrization and numerical fitting as described in section 4, both procedures are also automated. Moreover, the used supercontinuum laser enables to perform this experiment in a broad spectral range (400 – 2400 nm) that not only enables us to measure the spectral dependence of QMOKE but also serves as a credibility check for the deduced sample's magnetic anisotropy, which has to be spectrally-independent.

The functionality of the experimental setup and data processing was tested at temperature of 20 K using two 20-nm-thick $Ga_{1-x}Mn_xAs$ samples with nominal Mn dopings $x = 3\%$ and 7%. The deduced magnetic anisotropies were in accord with previously published results and, importantly, were the same for different wavelengths of detection light. QMOKE amplitudes were measured at several wavelengths and their anisotropies were evaluated. Interestingly, the QMOKE anisotropy was much smaller than that previously reported for Fe [8,15], Ni [23], $Fe_3O_4$ [45], Heusler compounds [25,46], and YIG [40]. This was attributed to a dominance of the (symmetric) kinetic exchange *p-d* coupling dominates over the (anisotropic) spin-orbit interaction in the Hamiltonian of (Ga,Mn)As.

As a last remark, we note that the eventual anisotropy of QMOKE amplitudes in a particular material should be seriously considered also in the QMOKE-based time-resolved pump-probe experiments [30,31,32], which are one of the few experimental tools available for the research of magnetic ordering in thin antiferromagnetic films [29].

**Acknowledgements**

The authors are indebted to V. Novák for the preparation of the studied samples. This work was supported by CzechNanoLab project LM2018110 funded by MEYS CR and also by TERAFIT project No. CZ.02.01.01/00/22_008/0004594 funded by OP JAK, call Excellent Research.

**Data availability statement**

Data reported in this paper has been deposited at Zenodo at doi: 10.5281/zenodo.14796045 and are publicly available as of the date of publication.

**Appendix A: General MO response of a thin film**

Here we derive the general form of the QMOKE response of a thin (001)-oriented ferromagnetic (Ga,Mn)As layer (zincblende crystal structure) on an optically isotropic substrate. For the purpose of this appendix, we strictly use the coordinate system of the cubic crystallographic axes (sample coordinate system: $x = [100], y = [010]$), see figure 1(b). Let us consider that the magnetization vector is localized in the sample plane $xy$ [50]. We describe the magnetization orientation by an inclination angle $\varphi_M$ with respect to the axis $x$. For crystals with the symmetry classes 432, $\bar{4}3m$ and $m3m$ [(Ga,Mn)As belongs to the $\bar{4}3m$ class], the relative permittivity tensor of the material can be written in the form [47]:

$$\epsilon = \begin{pmatrix} n_0^2 + \hat{G}_0 + \hat{G}_S \cos 2\varphi_M & 2\hat{G}_4 \sin 2\varphi_M & -\hat{K}\sin\varphi_M \\ 2\hat{G}_4 \sin 2\varphi_M & n_0^2 + \hat{G}_0 - \hat{G}_S \cos 2\varphi_M & \hat{K}\cos\varphi_M \\ \hat{K}\sin\varphi_M & -\hat{K}\cos\varphi_M & n_0^2 + \hat{G}_0 - \hat{G}_S \end{pmatrix}, \quad (A1)$$

where we defined the background (nonmagnetic) refractive index $n_0 = \sqrt{\varepsilon^{(0)}/\varepsilon_0}$ where $\varepsilon_0$ is the vacuum permittivity. The components of the linear magneto-optical tensor are $\hat{K} = K_{123} M_S/\varepsilon_0$ where $M_S$ is the saturated magnetization. We use the hat over the tensor components to discriminate our definition of rescaled components from the conventional one $K = K_{123}$. The quadratic tensor components are:

$$\hat{G}_0 = \frac{G_{1111} + G_{1122}}{2\varepsilon_0} M_S^2, \quad (A2)$$

$$\hat{G}_S = \frac{G_{1111} - G_{1122}}{2\varepsilon_0} M_S^2, \quad (A3)$$

$$\hat{G}_4 = \frac{G_{2323}}{2\varepsilon_0} M_S^2. \quad (A4)$$

The magneto-optical correction to the isotropic part of the permittivity is usually negligible ($\hat{G}_0 \ll n_0^2$) and therefore we neglect the $\hat{G}_0$ term in the following. Setting the angle of incidence of incoming linearly polarized light to 0° ensures zero linear MO response in our geometry due to the system symmetry and therefore QMOKE becomes the strongest MO effect. The terms of permittivity proportional to $\hat{K}$ still contribute, however, to the QMOKE [18]. On the other hand, our calculations show that their contribution is small ($\hat{K}^2/n_0^2 \ll \hat{G}_S, \hat{G}_4$) and therefore we neglect them here. In our geometry, the incident and consequently also both the reflected and transmitted electromagnetic fields are polarized in the sample plane $xy$. To find the eigenmodes $E_\pm$ of the electric field intensity inside the magnetic material, we solve the wave equation in the matrix form (omitting the z-component of the field as it is zero):

$$\left[\frac{\omega^2}{c^2}\begin{pmatrix} n_\pm^2 & 0 \\ 0 & n_\pm^2 \end{pmatrix} - \frac{\omega^2}{c^2}\begin{pmatrix} n_0^2 + \hat{G}_S \cos 2\varphi_M & 2\hat{G}_4 \sin 2\varphi_M \\ 2\hat{G}_4 \sin 2\varphi_M & n_0^2 - \hat{G}_S \cos 2\varphi_M \end{pmatrix}\right] E_\pm = 0. \quad (A5)$$

The symbol $\omega$ is the (monochromatic) field frequency, $c$ is the vacuum speed of light and $n_\pm$ are the effective refractive indexes of eigenmodes we are looking for. We find the solutions by setting the zero determinant of the above equation:

$$n_\pm^2 - n_0^2 = \pm\sqrt{\hat{G}_S^2 \cos^2 2\varphi_M + 4\hat{G}_4^2 \sin^2 2\varphi_M}, \quad (A6)$$

$$n_\pm \cong n_0 \pm \frac{\sqrt{\hat{G}_S^2 \cos^2 2\varphi_M + 4\hat{G}_4^2 \sin^2 2\varphi_M}}{2n_0} = n_0 \pm \delta n. \quad (A7)$$

In equation (A7), we have considered $n_\pm \cong n_0$ and we defined a small deviation from the non-magnetic refractive index $\delta n \ll 1$. Substituing back to equation (A5), the eigenmode electric intensities are found to be:

$$E_+ = N\left(2\hat{G}_4 \sin 2\varphi_M, \sqrt{\hat{G}_S^2 \cos^2 2\varphi_M + 4\hat{G}_4^2 \sin^2 2\varphi_M} - \hat{G}_S \cos 2\varphi_M\right), \quad (A8)$$

$$E_- = N\left(\sqrt{\hat{G}_S^2 \cos^2 2\varphi_M + 4\hat{G}_4^2 \sin^2 2\varphi_M} - \hat{G}_S \cos \varphi_M, -2\hat{G}_4 \sin 2\varphi_M\right), \quad (A9)$$

where $N$ is a normalization constant. It should be noted that the (complex) $G$ tensor components $\hat{G}_S$ and $2\hat{G}_4$ are not arbitrary. It is physically meaningful to require that the field eigenmodes are orthogonal for all $\bar{\varphi}_M$ such that $E_+^* \cdot E_- = 0$. This requirement directly and unambigously implies that both $\hat{G}_S$ and $2\hat{G}_4$ have the same complex phase: $\hat{G}_S^* \hat{G}_4 = \hat{G}_S \hat{G}_4^* \in \mathbb{R}$. The electric field components can be immediately rewritten as the (linearly polarized) vectors:

$$E_+ = \frac{1}{\sqrt{2Z(Z-C)}}(S, Z-C), \quad (A10)$$

$$E_- = \frac{1}{\sqrt{2Z(Z-C)}}(Z-C, -S), \quad (A11)$$

where we defined $S = 2\hat{G}_4(\hat{G}_4^*/|\hat{G}_4|) \sin 2\varphi_M = 2|\hat{G}_4| \sin 2\varphi_M$, $C = \hat{G}_S(\hat{G}_4^*/|\hat{G}_4|) \cos 2\varphi_M$ and $Z = \sqrt{C^2 + S^2}$. Now we make an estimate of the appropriate (Fresnel) reflection coefficients of the eigenmodes. It is possible since they are eigenmodes both of the sample material and its isotropic surroundings. Considering a semi-infinite sample and light incidence from vacuum, the reflection coefficients will be:

$$r_\pm = \frac{1-n_\pm}{1+n_\pm} = \frac{1-(n_0 \pm \delta n)}{1+(n_0 \pm \delta n)} \approx \frac{1-n_0}{1+n_0} \mp \frac{2\delta n}{(1+n_0)^2} = r_0 \mp \delta r. \quad (A12)$$

Here, we neglected the terms quadratic in $\delta n$ and we defined the non-magnetic reflection coefficient $r_0$ and its correction due to the QMOKE $\delta r$. For a thin sample, we must take into account also the reflection on the sample/substrate boundary and also multiple reflections and interferences. For the sake of the simplicity, we consider the sample to be a layer of thickness $d$ surrounded by vacuum from both sides; indeed, the result of the calculation is general, and it is valid for all isotropic substrates. The reflection coefficients of the eigenmodes from a thin layer read:

$$r_{layer\pm} = r_\pm \frac{1 - \exp[2i\frac{\omega}{c}n_\pm d]}{1 - r_\pm^2 \exp[2i\frac{\omega}{c}n_\pm d]} \ . \tag{A13}$$

First, we consider that the optical thickness of the layer is much less than the wavelength, in which case $2\frac{\omega}{c}n_\pm d \ll 1$ and $\exp[2i\frac{\omega}{c}n_\pm d] \cong 1 + 2i\frac{\omega}{c}n_\pm d$. The denominator can be therefore rewritten as

$$1 - r_\pm^2 \exp\left[2i\frac{\omega}{c}n_\pm d\right] \cong 1 - (r_0^2 \mp 2r_0\delta r)\left(1 + 2i\frac{\omega}{c}n_0 d\right) \tag{A14}$$

and the whole reflection coefficient can be written as:

$$r_{layer\pm} \cong -2i\frac{\omega}{c}\frac{n_0 r_0 \mp n_0\delta r \pm r_0\delta n}{1 - r_0^2\left(1 + 2i\frac{\omega}{c}n_0 d\right)}\left[1 \mp \delta r \frac{2r_0\left(1 + 2i\frac{\omega}{c}n_0 d\right)}{1 - r_0^2\left(1 + 2i\frac{\omega}{c}n_0 d\right)}\right] . \tag{A15}$$

In the first step, we used the approximations described above and in the second step, we used the expression $1/(1 - x) \cong 1 + x$ for $x \ll 1$. The result on the right-hand side can be further simplified by considering only the terms linear in $\delta n$ and thus also linear in $\delta r$. We may finally write $r_{layer\pm} = r_{layer 0} \mp \delta r_{layer}$ where, similarly to equation (A12), $\delta r_{layer} \propto (-\delta n)$ what is the key result here. Next, we continue with the calculation of the polarization rotation angle $\Delta\beta$. The reflected electric field intensity can be expressed as the linear combination of reflected eigenmodes:

$$\boldsymbol{E}_r = \boldsymbol{E}_+ r_+(\boldsymbol{E}_+^* \cdot \boldsymbol{E}_i) + \boldsymbol{E}_- r_-(\boldsymbol{E}_-^* \cdot \boldsymbol{E}_i) \tag{A16}$$

where $r_\pm$ are general reflection coefficients of the eigenmodes and $\boldsymbol{E}_i = E_i(\cos\beta, \sin\beta)$ is the incoming electric field vector. The complex conjugates of the vectors are left in the expression even though the vectors are real: all the expressions here are left to be generally applicable for the description of any MO effect, including the linear ones. Let us define the parallel and perpendicular vectors to the incoming polarization $\boldsymbol{E}_\parallel = (\cos\beta, \sin\beta)$ and $\boldsymbol{E}_\perp = (-\sin\beta, \cos\beta)$. The (small) polarization rotation $\Delta\beta(=\Delta\bar{\beta})$ upon reflection can be written then as:

$$\Delta\beta + i\epsilon \cong \frac{\boldsymbol{E}_\perp^* \cdot \boldsymbol{E}_r}{\boldsymbol{E}_\parallel^* \cdot \boldsymbol{E}_r} = \frac{(\boldsymbol{E}_\perp^* \cdot \boldsymbol{E}_+)(r_0 - \delta r)(\boldsymbol{E}_+^* \cdot \boldsymbol{E}_i) + (\boldsymbol{E}_\perp^* \cdot \boldsymbol{E}_-)(r_0 + \delta r)(\boldsymbol{E}_-^* \cdot \boldsymbol{E}_i)}{\left(\boldsymbol{E}_\parallel^* \cdot \boldsymbol{E}_+\right)(r_0 - \delta r)(\boldsymbol{E}_+^* \cdot \boldsymbol{E}_i) + \left(\boldsymbol{E}_\parallel^* \cdot \boldsymbol{E}_-\right)(r_0 + \delta r)(\boldsymbol{E}_-^* \cdot \boldsymbol{E}_i)} \tag{A17}$$

Here the variable $\epsilon$ is the ellipticity gained by the field. This result can be further simplified algebraically. The eigenmode vectors are an orthogonal basis of the vector space of dimension 2 and therefore $1 = \boldsymbol{E}_+\boldsymbol{E}_+^* + \boldsymbol{E}_-\boldsymbol{E}_-^*$, where the vector multiplication without the dot operator means the diadic product (resulting in a $2 \times 2$ matrix). For this reason, terms proportional to $r_0$ in the numerator sum up to $r_0\boldsymbol{E}_\perp^* \cdot (\boldsymbol{E}_+\boldsymbol{E}_+^* + \boldsymbol{E}_-\boldsymbol{E}_-^*) \cdot \boldsymbol{E}_i = r_0\boldsymbol{E}_\perp^* \cdot \boldsymbol{E}_i = 0$ and the same proportionality term in the denominator reads $r_0\boldsymbol{E}_\parallel^* \cdot (\boldsymbol{E}_+\boldsymbol{E}_+^* + \boldsymbol{E}_-\boldsymbol{E}_-^*) \cdot \boldsymbol{E}_i = r_0\boldsymbol{E}_\parallel^* \cdot \boldsymbol{E}_i = r_0 E_i$. The numerator is therefore proportional to the first power of $\delta r$ and so we can omit the term proportional to $\delta r$ in the denominator in order to get the linear dependency of $\Delta\beta + i\epsilon$ on $\delta r$ and consequently on $\delta n$. The general result is then:

$$\Delta\beta + i\epsilon = -\frac{\delta r}{r_0} \mathbf{E}_\perp^* \cdot (\mathbf{E}_+ \mathbf{E}_+^* - \mathbf{E}_- \mathbf{E}_-^*) \cdot \mathbf{E}_\parallel . \tag{A18}$$

With the help of equations (A10) and (A11), we may evaluate the above equation as:

$$\Delta\beta + i\epsilon = -\frac{\delta r}{Zr_0}(S \cos 2\beta - C \sin 2\beta) . \tag{A19}$$

Finally, we recognize with the help of equation (A7) that $\text{Re}[\delta r/r_0] \propto \delta n \propto Z$. We may therefore write the general result for the QMOKE, valid for both semi-infinite and thin samples on an isotropic substrate, and further considering the normal light incidence:

$$\Delta\beta = P_4 \sin 2\varphi_M \cos 2\beta - P_S \cos 2\varphi_M \sin 2\beta , \tag{A20}$$

where the coefficients $P_S \propto \hat{G}_S$ and $P_4 \propto 2\hat{G}_4$ are directly related (with a linear proportionality) to the quadratic magneto-optical tensor components and do not depend on neither $\beta$ nor $\varphi_M$. Their ratio is directly the ratio of the $G$ tensor components:

$$\frac{P_S}{P_4} = \frac{C}{S} = \frac{\hat{G}_S}{2\hat{G}_4} = \frac{G_{1111} - G_{1122}}{2G_{2323}} . \tag{A21}$$

In the special case of a fully isotropic medium, the QMOKE must not depend on the orientation of the axial system and therefore it must be a function of $\varphi_M - \beta$. We may then write $P = P_4 = P_S$ and:

$$\Delta\beta = P \sin 2(\varphi_M - \beta) = P \sin 2(\bar{\varphi}_M - \bar{\beta}), \tag{A22}$$

where the angles $\bar{\varphi}_M = \varphi_M + \chi$ and $\bar{\beta} = \beta + \chi$ are defined in the laboratory coordinate system such that they are rotated with respect to the crystallographic axes by the angle $\chi$ in the plane $xy$.

To quantify the MO amplitude, authors usually consider only s- and p-polarizations in literature [8,15]. This is mainly because they consider a non-zero angle of incidence in which case a derivation of a general formula for the QMOKE response would be too complicated and useless. In the case of the near-normal incidence, however, our approach is more general. Neglecting the $K$-tensor contribution, we can directly compare our results with the analysis reported previously: For example, equation (2) in [15] and equation (7) in [8] are special cases of our equation (3). Consequently, all used parameters can be converted between these equations. In particular, comparison with [8] and [25] yields $P_S = A_{s/p}G_S/2$ and $P_4 = A_{s/p}G_{44}$ (amplitudes of the s- and p-polarized responses $A_s$ and $A_p$ are equal at the normal incidence). It is also illustrative to express in our notation the dependency of the $M_L M_T$ and $M_L^2 - M_T^2$ terms, which appear in the eight-directional method [15,25], on the sample orientation angle $\chi$:

$$M_L M_T: \quad \Delta\beta = \frac{P_4 + P_S}{2} + \frac{P_4 - P_S}{2} \cos 4\chi , \tag{A23}$$

$$M_L^2 - M_T^2: \quad \Delta\beta = -\frac{P_4 - P_S}{2} \sin 4\chi . \tag{A24}$$

It is then possible to get the values $P_4$ and $P_S$ from the amplitudes of these dependencies and from the offset of the $M_L M_T$ contribution, see equation (A23). For example analysis of figure 3 of [25] yields $(P_4 + P_S)/2 = 8$ mdeg, $(P_4 - P_S)/2 = -20$ mdeg and thus $P_4 = -12$ mdeg and $P_S = 28$ mdeg. From these, the relative QMOKE anisotropy [equation (6)] of 21 nm Co$_2$FeSi [25] can be derived as $\delta P = +0.4$. The values of $|P_4|$, $|P_S|$ and $\delta P$ deduced from results available in literature are summarized in Table 2.

**Appendix B: Detailed data processing procedure**

In this appendix, we describe the procedure of an analysis of the full data set $\Delta\beta^{(\text{raw})}(\varphi_H, \beta)$ measured for polarization angles $\beta$ all around the full 360° circle. We aim at processing the measured data in order to be able to fit equation (3). The main difference between the experimental data and equation (3) is that the QMOKE polarization change is measured as a function of $\varphi_H$ instead of $\varphi_M$:

$$\Delta\beta(\varphi_H, \beta) = P_4 \sin 2\varphi_M(\varphi_H) \cos 2\beta - P_S \cos 2\varphi_M(\varphi_H) \sin 2\beta, \tag{B1}$$

where the dependency $\varphi_M(\varphi_H)$ is deterministic but unknown. An important fact here is, indeed, that equation (B1) has two terms where the dependencies on the two parameters $\varphi_H$ and $\beta$ are separated. For a fixed angle $\varphi_H$, the dependency $\Delta\beta(\beta)$ is therefore a linear combination of $\cos 2\beta$ and $\sin 2\beta$. Theoretically, fitting of the $\beta$-dependencies at each point $\varphi_H$ would allow us to get the individual curves $P_4 \sin 2\varphi_M(\varphi_H)$ and $P_S \cos 2\varphi_M(\varphi_H)$ from whose we would be able to directly get the QMOKE coefficients and the $\varphi_M(\varphi_H)$ relation, hence also the magnetic anisotropy. The experimental reality is more complex, however, and therefore we need to perform more steps as described below.

First, we do the symmetrizations of the measured raw data $\Delta\beta^{(\text{raw})}$ with respect to magnetic field and polarization, as described in section 4.1, in order to remove LinMOKE and a part of the experimental artifacts. By this procedure, we get the symmetrized data $\Delta\beta^{(\text{sym})}$. Second, we assume that all optical anisotropies in the experiment (including both the non-magnetic and the magneto-optically induced anisotropies of the sample and other optical components) are not much larger that the QMOKE-induced anisotropy plus we assume that the induced polarization changes are of the order of 10 mrad at most. Under these assumptions, all the polarization changes can be viewed as weak perturbations to the incoming field, their conjunctive effect is then negligible, and the resulting polarization modulation of the probe field is a sum of individual modulations. It is therefore important to achieve "as small angle of incidence of the probe field as possible" in order to minimize the LinMOKE and non-magnetic anisotropy caused by the nonzero angle of incidence.

Since the LinMOKE contributions are averaged out in the symmetrized data, rotation of the magnetic field results in the measurement of the curve $\Delta\beta^{(\text{sym})}(\varphi_H)$ for a fixed polarization angle $\beta$. This dependency results from QMOKE plus other eventual polarization rotations which do not depend on $\varphi_H$ and are therefore present as a constant offset. The reality is, however, that a combined effect of the imperfections of the polarization optics (in particular that of the balancing waveplate in the optical bridge) and the nonzero angle of incidence may lead to the appearance of terms proportional to $\sin 4\beta$ and $\cos 4\beta$. We checked this fact numerically and our analysis shows that these terms can be under some conditions comparable to the terms with the $2\beta$ argument and it is therefore important to consider and remove them from the measured data. Based on equation (B1) and the above discussion, we finally use the following ansatz for the symmetrized data:

$$\Delta\beta^{(\text{sym})}(\varphi_H, \beta) = \Delta\beta_4^{(\text{fit})}(\varphi_H) \cos 2\beta - \Delta\beta_S^{(\text{fit})}(\varphi_H) \sin 2\beta$$
$$+ S_1(\varphi_H) \cos 4\beta + S_2(\varphi_H) \sin 4\beta + D(\beta) + E(\varphi_H). \tag{B2}$$

The term $D(\beta)$ expresses contributions from the non-magnetic anisotropies and experimental uncertainty which arises due to an arbitrary choice of a reference point for balancing the waveplate in the optical bridge (see section 3 for more details). The term $E(\varphi_H)$ is the residuum which describes the eventual magnetic field dependency with unknown source, thus representing some error of the experimental setup. Importantly, equation (B2) does not express a unique fit: we may for example make the change $D \to D - \cos 2\beta$, $\Delta\beta_4^{(\text{fit})} \to \Delta\beta_4^{(\text{fit})} + 1$ and the right-hand-side of equation (B2) remains the same. We must therefore employ some constraint on $\Delta\beta_4^{(\text{fit})}(\varphi_H)$ and $\Delta\beta_S^{(\text{fit})}(\varphi_H)$ to eliminate the ambiguity. Comparing equations (B1) and (B2), we observe that we must require:

$$P_4 \sin 2\varphi_M(\varphi_H) = \Delta\beta_4^{(\text{fit})}(\varphi_H) , \tag{B3}$$

$$P_S \cos 2\varphi_M(\varphi_H) = \Delta\beta_S^{(\text{fit})}(\varphi_H) . \tag{B4}$$

The above equality is not exact due to experimental errors what means that equations (B3) and (B4) are valid in an ideal noiseless case. The squared left-hand-sides of equations (B3) and (B4), divided by $P_4^2$ and $P_S^2$, respectively, serve us the goniometric identity $\sin^2 \varphi_M + \cos^2 \varphi_M = 1$, and the searched constraint turns out to be:

$$\frac{1}{P_4^2}\left[\Delta\beta_4^{(\text{fit})}(\varphi_H)\right]^2 + \frac{1}{P_S^2}\left[\Delta\beta_S^{(\text{fit})}(\varphi_H)\right]^2 = 1. \tag{B5}$$

Taking the experimental noise back into consideration, equation (B5) is never met for all $\varphi_H$ and therefore the meaning of equation (B5) is, that we fit it with the smallest possible error. Equation (B5) does not indeed define the signs of the QMOKE coefficients $P_4$ and $P_S$ and therefore these must be set such that equations (B3) and (B4) give the correct form of equation (B1).

The first approach in determining the magnetic anisotropy is the direct reconstruction of the relation $\varphi_M(\varphi_H)$ from equations (B3) and (B4):

$$\frac{P_S}{P_4} \frac{\Delta\beta_4^{(\text{fit})}(\varphi_H)}{\Delta\beta_S^{(\text{fit})}(\varphi_H)} = \tan 2\varphi_M(\varphi_H). \tag{B6}$$

The second possibility, how to implement equations (B3) and (B4), is to consider explicitly that the magnetization is the subject of an in-plane quadratic and uniaxial anisotropies (for our GaAs-based cubic material, due to the symmetry considerations appropriate for the particular sample) as described by equation (7). We then fit the parameters $H_u$, $H_c$ and $\varphi_u$ to get best agreement with equations (B3) and (B4).

# Supplementary material:

# Quadratic magneto-optical Kerr effect spectroscopy: Polarization variation method for investigation of magnetic and magneto-optical anisotropies

V. Wohlrath,[1] Z. Sadeghi,[1] J. Kimák,[1] K. Hovořáková,[1] P. Kubaščík,[1] E. Schmoranzerová[1], L. Nádvorník,[1] F. Trojánek,[1] P. Němec,[1,*] and T. Ostatnický[1]

[1]*Faculty of Mathematics and Physics, Charles University,*
*Ke Karlovu 3, 121 16 Prague 2, Czech Republic*

*Author to whom any correspondence should be addressed.
E-mail: petr.nemec@matfyz.cuni.cz.

## CONTENTS

## S1. Construction of vector electromagnet

When constructing an electromagnet for our MO experiment, we were inspired by [S1]. However, as we aimed for a full-360° rotation in a 2D plane using a cryostat and magnetic fields of at least 200 mT, we had to construct a much bigger electromagnet. It consists of two independent pairs of coils with coils in a pair connected serially [see figure S1(a)]. Here, a magnetic field is generated by currents $I_1$ and $I_2$ flowing through coils. The currents are provided by two independent current supplies (BOP 50-20GL made by Kepco). Four poles of electromagnet, with a spacing of 65 mm, redirect the magnetic field into a sample space, where the cryostat is located [sample is depicted as the green square in figure S1(a)]. The used laboratory coordinate system is depicted in figures S1(a) and S1(b)]. For a comparison of the laboratory and the sample coordinate systems see figure 1 in the main text.

For our MO measurements, we need to know not only the dependency of the magnetic field magnitude on the currents $I_1$ and $I_2$ in a single point where the sample is placed (axes origin) but also its spatial distribution, namely $\mu_0 H(\bar{x}, \bar{y})$ and $\bar{\varphi}_H(\bar{x}, \bar{y})$. To check the magnetic field homogeneity, we used a custom-made 3D Hall probe [figure S1(c)], which has three Hall sensors [figure S1(d)] on it. This 3D Hall probe is able to measure projections of the magnetic field onto three perpendicular axes. Importantly, as the centers of the three Hall sensors do not lie in the same spatial point, the measurements of spatially displaced 3D Hall probe must be combined to determine the magnetic field amplitude $\mu_0 H$ and direction $\bar{\varphi}_H$ in a single point in space. Moreover, due to the symmetry of our electromagnet and a small film thickness in the studied samples, we are only interested in the 2D spatial dependence of the magnetic field in the $\bar{x}\bar{y}$ plane.

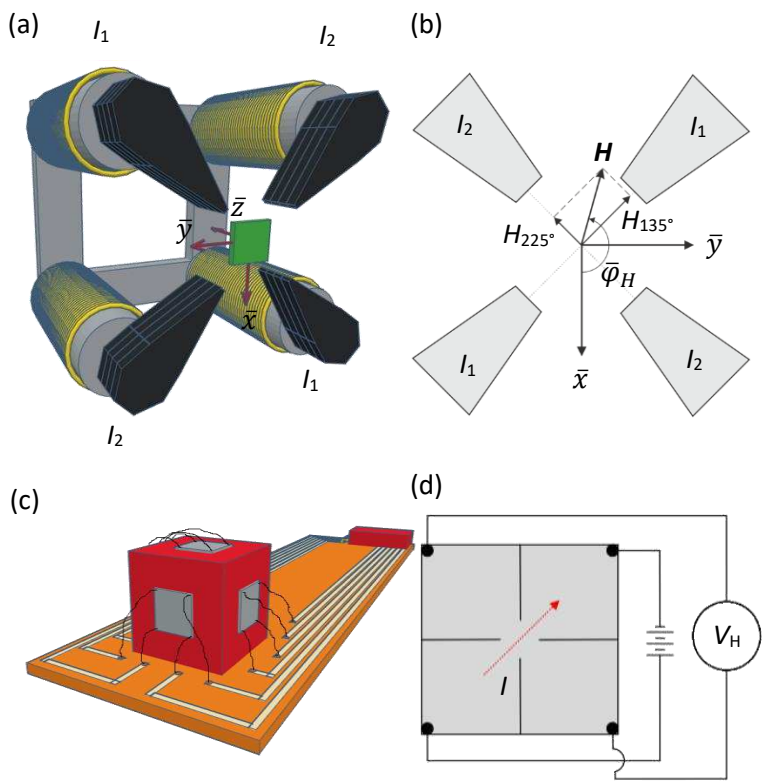

**Figure S1.** (a) Schematic depiction of the four-pole electromagnet generating magnetic field, whose amplitude and direction can be changed within the $\bar{x}\bar{y}$ plane. Light is incident along $+z$ direction and reflected along $-z$ direction. Coils in a single pair are connected serially, but two pairs of coils $I_1$ and $I_2$ are independent of each other. The green square represents a sample, laying in the $\bar{z} = 0$ plane. (b) Coordinate system and its position within the poles of the electromagnet. $H_{135°}$ and $H_{225°}$ represent projections of the magnetic field intensity vector $H$ onto the axes of the $I_1$ and $I_2$ poles, respectively. (c) Schematic drawing of the 3D Hall probe consisting of three independent Hall sensors. (d) Wiring diagram of a single Hall sensor. Red arrow indicates the direction of current $I$. $V_H$ is the Hall voltage.

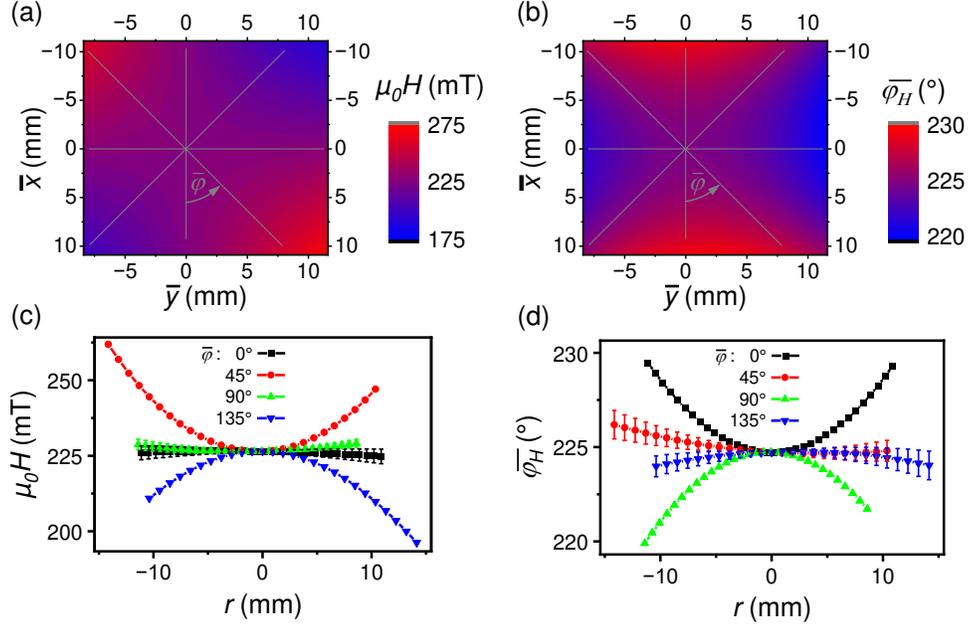

**Figure S2.** Example of magnetic field homogeneity. Spatial dependence of (a) amplitude and (b) direction of magnetic field within the $\bar{z} = 0$ plane for currents $I_1 = 0, I_2 = 20$ A. Angle $\bar{\varphi}$ represents directions of cuts (depicted as grey lines) through the coordinate system origin. Dependence of (c) amplitude and (d) direction of magnetic field as a function of distance $r$ from the coordinate system origin for selected $\bar{\varphi}$.

Example of the measured spatial dependence of the magnetic field for $I_1 = 0, I_2 = 20$ A is shown in figures S2(a) and S2(b). As we are applying voltage only on coils $I_2$ in this case, the generated magnetic field is mostly along the $\bar{\varphi}_H = 225°$ direction in the whole $\bar{z} = 0$ plane, as depicted in figure S2(b). Moreover, the closer we are to the $I_2$ poles, the bigger the field is, see figure S2(a). These data and data from other scans allowed us to determine that in the area $5 \times 5$ mm$^2$ around the origin, the magnetic field amplitude changes as little as ~2% and the magnetic field direction changes no more than ~1.5°. This area is large enough for our MO measurements, since the laser beam spot area in our experimental setup is considerably smaller than 1 mm$^2$.

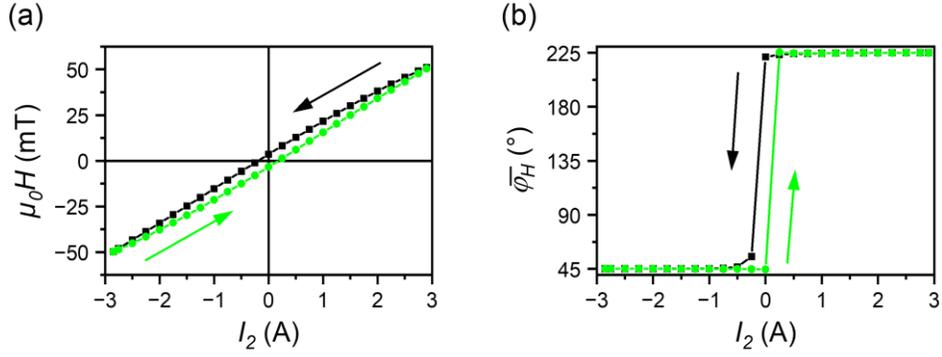

**Figure S3.** Magnetic field generated by a single pair of coils. Dependence of (a) amplitude and (b) direction of magnetic field on current flowing through a single pair of coils. Negative values correspond to reverse polarities. Arrows indicate the current change direction.

## S2. Driving of electromagnet

Changing magnetic field direction with no amplitude modulation, or vice versa, is often needed during MO experiments. Both of these can be generated by our four-pole electromagnet (see section S1 of this Supplementary material), however, magnetic characteristics of materials used in the electromagnet construction must be taken into account.

Ferromagnetic materials display hysteretic behavior – magnetic field generated by a ferromagnet depends on both external magnetic field and its history. In our case, magnetic fields are created by the currents $I_1$ and $I_2$. Therefore, we show $\mu_0 H(I_j)$ and $\bar{\varphi}_H(I_j)$ dependences in our graphs rather than the usual $M(H)$. (Note that we are always showing the magnetic field in the coordinate system origin, where the sample is placed.) Here, $\mu_0 H$ and $\bar{\varphi}_H$ are magnetic field amplitude and direction, respectively, created by the electromagnet and $I_j$ is the current though coils $I_1$ or $I_2$. An example of such dependencies is shown in figure S3 where the hysteretic behavior is clearly apparent. When creating a sequence of currents $I_1, I_2$, which would generate a desired sequence of magnetic fields $\mu_0 \mathbf{H}$, one must therefore consider the measurement ranges for both currents $I_1, I_2$ and the appropriate branches of the hysteresis loops.

To illustrate the importance of choosing the correct branch of the hysteresis loop, we show the difference between rotating magnetic field in a clockwise and a counter-clockwise sense around the direction $\bar{\varphi}_H = 45°$ in figure S4. An important note: currents $I_1$ and $I_2$ with the lowest (negative) values generate magnetic fields in the directions 315° and 45°, respectively [see figure S1(b)].

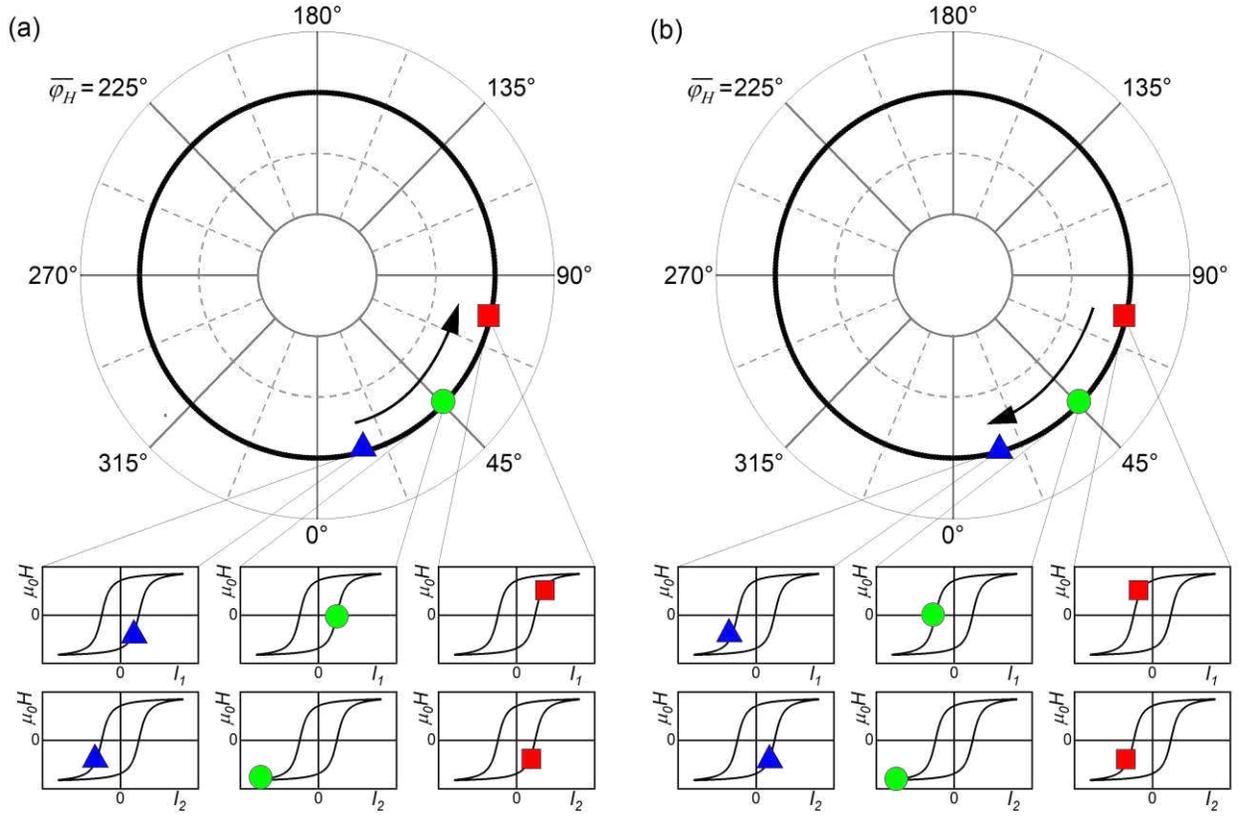

**Figure S4.** Sketch of magnetic field magnitude and orientation in the polar plot when rotating the magnetic field direction around a point where the magnetic field amplitude generated by a single pair of coils does not change monotonously with current. Changing the magnetic field direction $\bar{\varphi}_H$ from (a) the blue point to the red one in the clockwise sense and from (b) the red point to the blue one in the counter-clockwise sense. Black arrows indicate the change of the magnetic field direction.

In figure S4(a), the desired counter-clockwise change of the magnetic field direction from the blue point to the red one requires a monotonous increase of $\mu_0 H_{135°}$, *i.e.*, a projection of the magnetic field vector to the axis of the $I_1$ poles [*cf.* figure S1(b)]. The increase of $\mu_0 H_{135°}$ means choosing the lower branch of the hysteresis loop $\mu_0 H(I_1)$. However, the projection $\mu_0 H_{225°}$ of the magnetic field vector must first go down to its minimum through the upper branch of the hysteresis loop $\mu_0 H(I_2)$ and then raise up through the lower branch. Because the shape of the hysteresis loop depends on the applied current extreme, skipping the middle step [green point in figure S4(a)] would lead to a different (minor) hysteresis loop of $\mu_0 H(I_2)$.

The same approach holds true for the clockwise change of the magnetic field direction [figure S4(b)]. This case requires a monotonous decrease of $\mu_0 H_{135°}$, therefore choosing the upper branch of the hysteresis loop $\mu_0 H(I_1)$ and exactly the same change of $\mu_0 H_{225°}$ as in the previous case.

With this approach, we were able to obtain rotational magnetic field with amplitudes $\mu_0 H = (207 \pm 5)$ mT, depicted in Fig S5(a), and $\mu_0 H = (50 \pm 2)$ mT, figure S5(b), both with the angular step of $\sim 1°$. We also created sequences of magnetic fields, where only the amplitude changes, as illustrated in figure S6.

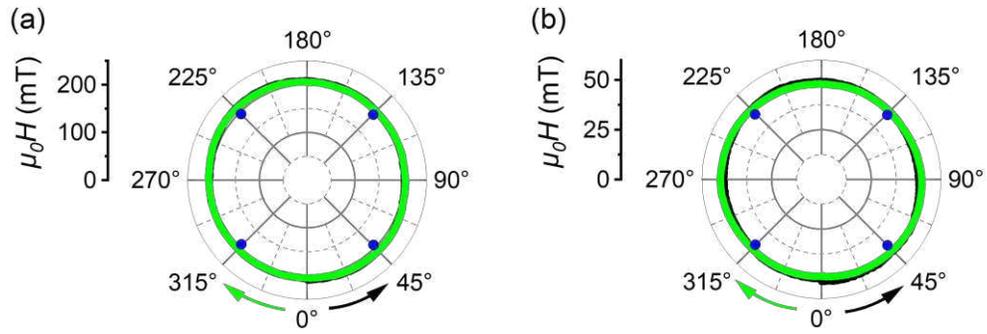

**Figure S5.** Angular dependence of the magnetic field amplitude when changing its direction for (a) 207 mT and (b) 50 mT. Blue dots represent points the electromagnet must go through when making a full 360° turn. Arrows indicate the magnetic field direction change.

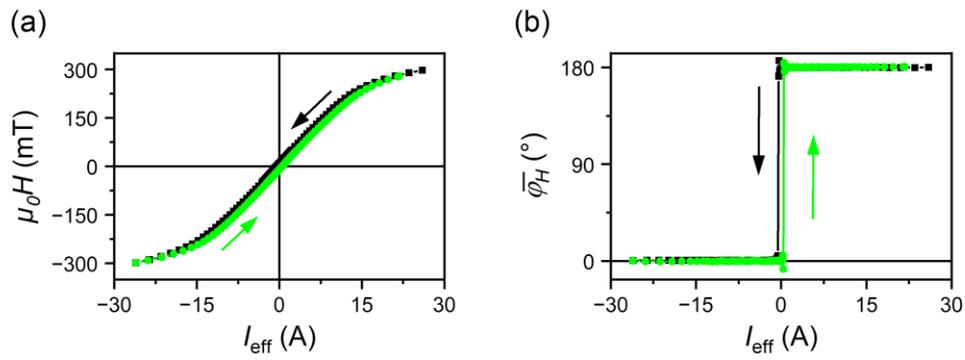

**Figure S6.** Change of magnetic field amplitude without changing its direction. (a) Amplitude and (b) direction of the magnetic field as functions of the effective current defined as $I_{\text{eff}} = \text{sign}(I_2) \cdot \sqrt{I_1^2 + I_2^2}$. Arrows indicate the effective current change.

## S3. Discussion of numerical fitting method for deducing magnetic anisotropy from measured MO data

In this section, we provide a detailed discussion of the MO data fitting, according to section 4.3 in the main text. The general assumption throughout the used procedure is that the QMOKE response is caused solely by the sample while other (paramagnetic) elements in the setup reveal only LinMOKE. Therefore, the MO response of all components but the sample is removed by the symmetrization with respect to the magnetic field as discussed in section 4.1 in the main text. The symmetrization is demonstrated in figure 3(a). Clearly, the raw and (magnetic field-) symmetrized data differ, confirming the importance of this step. The additional symmetrization in probe polarization should theoretically have no major impact on the data: indeed, as seen in the figure, the correction in this particular case is of the order of several per cents only. We interpret this fact as a consequence of a relatively small imperfection of the used high-quality polarization optics. On the other hand, our numerical analysis (not shown here) confirms that the symmetrization in polarization is capable of removing a dominant portion of experimental errors even for lower quality polarization optics and the symmetrized data are therefore more reliable for a determination of the QMOKE coefficients.

The symmetrization in polarization removes only a part of the experimental error, however. We illustrate this fact in figure 5(a) in the main text: it is worth comparing two symmetrized curves $\pm \Delta\beta^{(\text{sym})}(\varphi_H, \beta = \pm 45°)$ which should theoretically overlap, according to equation (3). In reality, these two curves differ, as seen in figure 5(a), revealing that the symmetrization has left the error in a form of not only a random noise but rather some function of both $\beta$ and $\varphi_H$, see also section S3. We ascribe this weak contribution to some effect of the nonzero angle of incidence of the probe light and we justify our fitting function equation (B2) in the following.

Let us first consider a fixed angle $\varphi_H$ of the magnetic field. The lowest-order term for the QMOKE is small ($\Delta\beta \ll 1$) and equation (A20) predicts its proportionality to a harmonic function of $2\beta$. The rotation induced by the nonzero angle of incidence only (neglecting any MOKE here) is calculated similarly, using equation (A18). We substitute the $s$- (index $s$) and $p$-polarization (index $p$) vectors for $\boldsymbol{E}_\pm$, i.e. $\boldsymbol{E}_+ = \boldsymbol{E}_s = (1,0)$, $\boldsymbol{E}_- = \boldsymbol{E}_p = (0,1)$, assuming the plane of incidence $yz$. The $z$-component of the $p$ wave can be neglected here due to the small angle of incidence. The reflection coefficients become $r_0 = (r_s + r_p)/2$ and $\delta r = (r_p - r_s)/2$. Finally, we get the change of the polarization state:

$$\Delta\beta + i\epsilon = 2\frac{r_p - r_s}{r_p + r_s}\sin 2\beta. \tag{S1}$$

Note that $|\Delta\beta + i\epsilon| \ll 1$ since $r_p \approx r_s$ for small angles of incidence. Similarly to the QMOKE, this result is a harmonic function of $2\beta$ but it does not depend on $\varphi_H$ and cannot therefore by itself explain the noticeable differences between the aforementioned curves in figure 5(a). Considering both the QMOKE and nonzero angle of incidence, it turns out that there exists some weak, yet detectable effect which has the symmetry of the product of the two: a constant in $\beta$ plus a harmonic function of $4\beta$, both depending also on $\varphi_H$. This is where the terms $S_{1,2}(\varphi_H)$ and $E(\varphi_H)$ in equation (B2) come from. In particular, $\Delta\beta^{(\text{sym})}(\beta = 45°)$ is then $-\Delta\beta_S^{(\text{sym})} + S_2$ while we get $-\Delta\beta^{(\text{sym})}(\beta = -45°) \sim -\Delta\beta_S^{(\text{sym})} - S_2$. The nonzero difference $2S_2$ between $\Delta\beta^{(\text{sym})}(\beta = +45°)$ and $\Delta\beta^{(\text{sym})}(\beta = -45°)$ is therefore what we expect. This difference is furthermore a function of $\varphi_H$ because also $S_2$ is the function of $\varphi_H$, in accordance with the data in figure 5(a).

The above discussion leads unambiguously to the conclusion that the symmetrized data contain an additional $\varphi_H$-dependent contribution as compared to equation (3) and therefore equation (10) can result in relatively erroneous $\varphi_M(\varphi_H)$ dependency when we use symmetrized data as its input. Nevertheless, the symmetrization is sufficient for determination of the QMOKE coefficients and, consequently, determination of the ratio between the MO permittivity tensor elements as shown in equation (A21).

To correctly determine the magnetic anisotropy, on the other hand, one needs to get rid of the contributions denoted as the $S_{1,2}$ and $E$ terms in equation (B2). The discussion above implies that the average $[\Delta\beta^{(\text{sym})}(\beta) - \Delta\beta^{(\text{sym})}(\beta + 90°)]/2$ removes these artifacts and thus we can make estimates:

$$\Delta\beta_4^{(\text{fit})}(\varphi_H) \approx \frac{1}{2}[\Delta\beta^{(\text{sym})}(\varphi_H, \beta = 0°) - \Delta\beta^{(\text{sym})}(\varphi_H, \beta = 90°)], \tag{S2}$$

$$\Delta\beta_S^{(\text{fit})}(\varphi_H) \approx -\frac{1}{2}[\Delta\beta^{(\text{sym})}(\varphi_H, \beta = +45°) - \Delta\beta^{(\text{sym})}(\varphi_H, \beta = -45°)]. \tag{S3}$$

These estimates then enter equations (B3), (B4). To do so, we need to measure at least eight not four incoming polarizations (0°, 45°, 180° and 225°), which are used in section 4.2 for deducing the QMOKE anisotropy. The better way than to use equations (S2) and (S3) is, in our opinion, to fit the symmetrized data by equation (B2). Doing so, it is possible to numerically correct the eventual small misalignments in the experimental setup due to small unintentional shifts of the angles $\beta$ and $\varphi_H$.

To gain a better insight into the full data analysis as described in section 4.3 and to estimate the error of the symmetrized data, we plot in figure S7 the curves which come from the processing of the measured MO data, the example of which is shown in figure 4(b) for the sample $x = 3\%$. We compare $\Delta\beta_S^{(\text{fit})}$ which contains information for further evaluation of the anisotropies, $S_2$ is proportional to the artificial $4\beta$ term and finally the fitting residuum $E(\varphi_H)$ is an error with an unknown source. Quite interestingly, the $E(\varphi_H)$ term is almost the same for both samples, which were measured at different wavelengths. Consequently, it seems to be that it is related solely to the used experimental setup, and it is sample and wavelength independent. Its amplitude, therefore, can be used to estimate the error of the QMOKE coefficients determination by fitting as 0.03 mrad. The method which uses the data symmetrization only for the QMOKE coefficients evaluation, on the other hand, does not remove the $S_2$ term and, therefore, we must add the amplitude of the $S_2$ curve to its error, as shown in table 1.

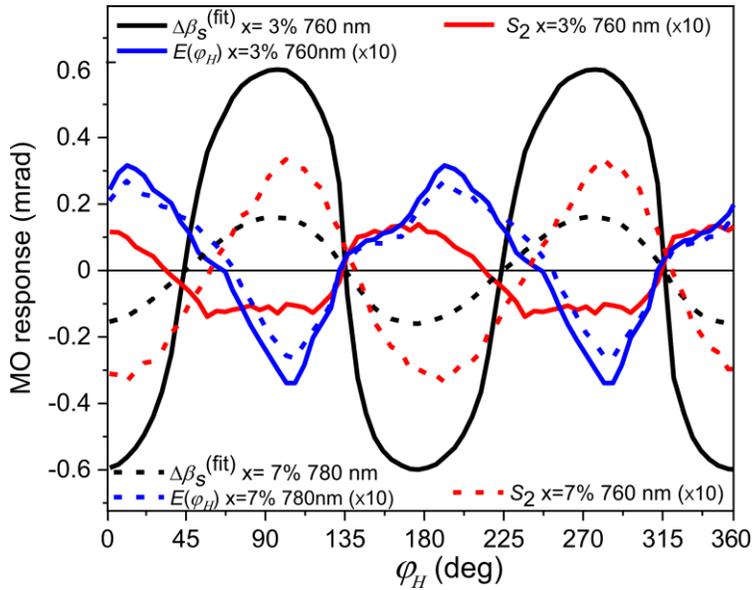

**Figure S7.** Selected fitted functions from equation (B2) for samples $x = 3\%$ and $x = 7\%$ at wavelengths of 760 nm and 780 nm, respectively.